\pdfoutput=1
\documentclass[aps,preprint,showpacs]{revtex4}%
\usepackage{amssymb}
\usepackage{amsfonts}
\usepackage{amsmath}
\usepackage{graphicx}%
\providecommand{\U}[1]{\protect\rule{.1in}{.1in}}

\begin{document}
\title[Quasibound states]{A Unified Theory of Quasibound States}
\author{Curt A. Moyer}
\affiliation{Department of Physics and Physical Oceanography, UNC Wilmington}
\email{moyerc@uncw.edu}

\begin{abstract}
We have developed a formalism that includes both quasibound states with real
energies and quantum resonances within the same theoretical framework, and
that admits a clean and unambiguous distinction between these states and the
states of the embedding continuum. States described broadly as `quasibound'
are defined as having a connectedness (in the mathematical sense) to true
bound states through the growth of some parameter. The approach taken here
builds on our earlier work by clarifying several crucial points and extending
the formalism to encompass a variety of continuous spectra, including those
with degenerate energy levels. The result is a comprehensive framework for the
study of quasibound states. The theory is illustrated by examining several
cases pertinent to applications widely discussed in the literature.

\end{abstract}
\date{June 24, 2013}

\maketitle

\section{Introduction}

The long history of quasibound states dates back to the earliest days of
quantum mechanics with Schr\"{o}dinger's successful calculation of Stark
splittings in atomic hydrogen \cite{Schrodinger}. Only later was it realized
that the Stark spectrum had no true eigenvalues (a continuous spectrum), and
that the perturbation series used to calculate the Stark levels is, in fact,
[asymptotically] divergent! The physical mechanism giving rise to this state
of affairs ultimately was traced to the existence of quasi-stationary states
caused by the perturbation of true bound states. The unstable nature of these
states was recognized by Oppenheimer, who apparently was the first to estimate
their lifetime \cite{Oppenheimer}. A similar idea was subsequently advanced by
Gamow \cite{Gamow} to explain alpha emission from radioactive nuclei, and by
Fowler and Nordheim \cite{Fowler} to account for the `cold' emission of
electrons from metals. Subsequently, quasi-stationary states came to be known
as resonances or later, quasibound states: all refer to states with a high
degree of localization embedded in an energy continuum of de-localized states
\cite{UsageNote}. The excellent review by Harrell \cite{Harrell} traces the
development of the mathematical foundations of quantum resonance theory from
its earliest days up to the present.

Recently, there has been a resurgence of interest in quasibound states, driven
in large part by their newly-discovered relevance to the areas of nanoscience
and electronic devices. Semiconductor quantum wells (QWs) and superlattices
(SLs), which have been widely studied for more than three decades, have
interesting electronic and optoelectronic properties that arise from the
localization of carrier wave functions. Accurate knowledge of the energy
levels and wave functions of electrons in these structures is vital to their
design, and resonant phenomena have been shown to be critical, both in
understanding their principles of operation and in boosting performance. High
performance QWIPs (Quantum Well Infrared Photodetectors) for example, rely on
bound-to-quasibound rather than bound-to-bound transitions to enhance their
sensitivity \cite{Gunapala}. And the kinetics of quasibound states have been
identified as playing an essential role in the performance of quantum cascade
lasers \cite{Razeghi}. Lastly, recent photoreflectance experiments \cite{Lee}
reveal the signature of quasibound states localized in a thick GaAs barrier of
(In,Ga)As/GaAs heterostructures having AlAs interface layers.

The general theory of resonances in nanostructures was advanced by Price
\cite{Price}, and efficient methods for their calculation soon followed
\cite{Anemogiannis}-\cite{Rihani}. The methods fall into two broad categories,
but each has its drawbacks. The simpler (from a conceptual standpoint) methods
are based on an identification of quasibound state energies with peaks in the
continuum density of states (DOS). The width of the DOS peaks is taken as a
measure of state lifetime, but the relation is imprecise and the widths
themselves can be difficult to calculate for broad or overlapping resonances.
Aside from their relative ease of computation, the DOS-based methods possess
another virtue: they yield \textit{real} quasibound state energies with
well-behaved wave functions that are essential for calculating related
quantitites of interest, e.g., dipole matrix elements involving quasibound
states. The alternative methods rely on the rigorous definition of resonances
as singularities of the scattering matrix ($S$-matrix), and yield
complex-valued energies whose imaginary part prescribes unambiguously the
lifetime of these metastable states. However, the wave functions for these
complex-energy states increase exponentially at the ends of the quantum
structure, and are unbounded at infinity. Such behaviour is unphysical, and
dipole matrix elements involving these exponentially-increasing waveforms are ill-defined.

As early as 1997, we championed another approach to characterizing quasibound
states \cite{Moyer} which has been largely overlooked in semiconductor device
studies, perhaps because it was too narrow in scope (quasibound states in a
uniform electric field). The quasibound states described there have real
energies with physically acceptable wave functions, thereby sharing the
virtues of the DOS-based methods described above. This feature was exploited
effectively in Ref.\cite{Moyer} to study the behavior of the induced electric
dipole moment in fields of arbitrary strength. The theory presented here
builds on that earlier work by clarifying several crucial points and extending
it to encompass a variety of continuous spectra, including those with
degenerate energy levels. The result is a comprehensive framework for the
study of quasibound states. Central to these developments is the introduction
of a new basis -- essentially a Fourier time transform of the spectral basis
-- that we have dubbed the \textit{quantum history}.

The plan of the paper is as follows: Following a brief introduction to quantum
histories in Sec. II, a general method for selecting quasibound states is
proposed and elaborated in Secs. III and IV. Of necessity, this demands a
precise definition of the term `quasibound'; the one we adopt is rooted in a
connectedness between the quasibound state and a true bound state brought
about by the variation of some physical parameter. In Sec. V we argue that
this definition combined with the requirement of gauge invariance leads to
just two fundamental types of quasibound states, viz., (1) quantum resonances,
and (2), true stationary states, with real energies and bounded (though not
square-integrable) wave functions. Secs. VI--VIII are devoted to developing
concrete methods for calculating these stationary quasibound states in several
important applications, distinguished by the spectral characteristics
(boundedness, degeneracy) of the embedding continuum. The principal results
are summarized in Sec. IX. Throughout we adopt natural units in which
$\hbar=1$, a choice that leads to improved transparency by simplifing numerous expressions.

\section{Quantum Histories: A Novel Basis Set}

For a quantum system described by the Hamiltonian operator $\widehat{H}$, we
define a set of states $|\,\tau\,\rangle$ in the Hilbert space to have the
property%
\begin{equation}
|\,\tau+\tau^{\prime}\,\rangle=\exp\left(  -i\widehat{H}\tau\right)
|\,\tau^{\prime}\,\rangle\label{basis property}%
\end{equation}
The states in Eq.(\ref{basis property}) are labeled by a continuous variable
$\tau$ that we call the \textit{system time}; it extends over the whole domain
of reals from the remote past ($\tau=-\infty$) to the distant future
($\tau=+\infty$). We refer to the set $\left\{  |\,\tau\,\rangle:-\infty
\leq\tau\leq\infty\right\}  $ as a \textit{timeline }or \textit{quantum
history}; it is distinguished by the requirement that its elements (the
\textit{time states} $|\,\tau\,\rangle$) can be combined to represent any
physical state. We advance the conjecture that \textit{a timeline exists for
every system, and constitutes a complete basis in the Hilbert space}. The
completeness of this basis is expressed by the closure rule for time states;
formally,%
\begin{equation}
1=%
{\displaystyle\int\limits_{-\infty}^{\infty}}
|\,\tau\,\rangle\langle\,\tau\,|\,d\tau\label{closure rule}%
\end{equation}
While the time states are complete, they are not usually orthogonal. Quantum
histories are fascinating in their own right, and have been addressed at
length in a separate publication \cite{Moyer4}; here we are content with
presenting only those features essential to the task at hand.

Timelines are intimately related to spectral structure, and derivable from it.
Indeed, the eigenstates of $H$, say $|\,E\,\rangle$, also span the space of
physically realizable states to form the \textit{spectral basis} $\left\{
|\,E\,\rangle:\forall E\right\}  $. According to Eq.(\ref{basis property})
(with $\tau^{\prime}=0$), the transformation from a timeline to the spectral
basis is characterized by functions $\langle\,\tau\,|\,E\,\rangle$ with the
property%
\begin{equation}
\langle\,\tau\,|\,E\,\rangle=const\cdot\exp\left(  iE\tau\right)  ,
\label{timeline-spectral link}%
\end{equation}
where $const=\langle\,\tau\,|\,E\,\rangle_{\tau=0}$ may still be a function of
energy $E$. In fact, Ref.\cite{Moyer4} establishes that $const$ must be
independent of $E$ to satisfy closure, and goes on to show how timelines can
be constructed for virtually any system, starting from the stationary states
of the associated Hamiltonian. The role of degeneracy deserves special
comment: \textit{a spectrum with }$n$\textit{-fold degenerate levels breeds
}$n$\textit{ distinct histories}. In other words, with degeneracy we get not
just one, but multiple quantum histories with mutually orthogonal timelines,
all of which must be included to span the Hilbert space of physically
realizable states.

\section{Identifying Quasibound States}

We seek to identify quasibound states described by the eigenvalue problem
\begin{equation}
\left(  \widehat{H}_{0}+\widehat{V}\right)  \,|\,\psi_{E}\,\rangle
=E\,|\,\psi_{E}\,\rangle\label{eigenvalue problem}%
\end{equation}
$\widehat{H}_{0}$ is that piece of the Hamiltonian giving rise to the
embedding continuum; it is time-independent and Hermitian, and normally would
include -- but is not limited to -- the kinetic energy. For the present
discussion we assume that the spectrum of $\widehat{H}_{0}$ is non-degenerate,
with an associated single quantum history $\left\{  |\,\tau\,\rangle
:-\infty\leq\tau\leq\infty\right\}  $ (this restriction will be lifted later).
$\widehat{V}$ is the interaction that is the source of the quasibound states,
and is assumed to have \textit{compact support}, so that the stationary state
waveforms in the asymptotic region take the same mathematical form with or
without $\widehat{V}$. Thus, the spectrum of $\widehat{H}_{0}+\widehat{V}$
also is continuous. To select the quasibound states from this continuum, we
will formally solve Eq.(\ref{eigenvalue problem}) in the basis that is the
quantum history of $\widehat{H}_{0}$.

Appealing to the infinitesimal version of Eq.(\ref{basis property})%
\[
\widehat{H}_{0}|\,\tau\,\rangle=i\frac{d}{d\tau}|\,\tau\,\rangle
\]
we obtain from Eq.(\ref{eigenvalue problem}) an integro-differential equation
for the timeline wavefunction $\psi_{E}(\tau)\equiv\langle\,\tau\,|\,\psi
_{E}\,\rangle$:%
\[
\left(  E+i\frac{d}{d\tau}\right)  \psi_{E}(\tau)=%
{\displaystyle\int_{-\infty}^{\infty}}
\langle\,\tau\,|\widehat{V}|\,\tau^{\prime}\,\rangle\psi_{E}(\tau^{\prime
})\,d\tau^{\prime}%
\]
With the help of the integrating factor $\exp\left(  -iE\tau\right)  $, we
formally integrate this over the interval $\left(  \tau_{0},\tau\right)  $ to
obtain an integral equation for $\psi_{E}(\tau)$
\begin{equation}
\psi_{E}(\tau)=\exp\left(  iE\tau\right)  \exp\left(  -iE\tau_{0}\right)
\psi_{E}(\tau_{0})+%
{\displaystyle\int_{-\infty}^{\infty}}
\left[
{\displaystyle\int_{-\infty}^{\infty}}
G_{E}\left(  \tau,\tau^{\prime\prime};\tau_{0}\right)  \langle\tau
^{\prime\prime}\,|\widehat{V}|\,\tau^{\prime}\rangle\,d\tau^{\prime\prime
}\right]  \psi_{E}(\tau^{\prime})\,d\tau^{\prime}, \label{integral equation}%
\end{equation}
where $G_{E}\left(  \tau,\tau^{\prime\prime};\tau_{0}\right)  $ is a
\textit{Green's function} with parameters $E$ and $\tau_{0}$, and defined by
the rules%
\begin{equation}
G_{E}\left(  \tau,\tau^{\prime\prime};\tau_{0}\right)  =\text{ \ }%
\begin{array}
[c]{ccc}%
-i\exp\left[  iE\left(  \tau-\tau^{\prime\prime}\right)  \right]  & \qquad &
\tau_{0}\leq\tau^{\prime\prime}\leq\tau\\
+i\exp\left[  iE\left(  \tau-\tau^{\prime\prime}\right)  \right]  & \qquad &
\tau\leq\tau^{\prime\prime}\leq\tau_{0}\\
0 & \qquad & \text{otherwise}%
\end{array}
\text{ \ } \label{Green's function}%
\end{equation}

To recover the abstract (Hilbert space) version of Eq.(\ref{integral equation}%
), we regard the function $G_{E}\left(  \tau,\tau^{\prime\prime};\tau
_{0}\right)  $ as timeline matrix elements of a Green operator $\widehat{G}%
_{E}\left(  \tau_{0}\right)  $, writing $G_{E}\left(  \tau,\tau^{\prime\prime
};\tau_{0}\right)  \equiv\langle\,\tau\,|\widehat{G}_{E}\left(  \tau
_{0}\right)  |\,\tau^{\prime\prime}\,\rangle$. Then the integral term on the
right of Eq.(\ref{integral equation}) becomes simply $\langle\,\tau
\,|\widehat{G}_{E}\left(  \tau_{0}\right)  \widehat{V}|\,\psi_{E}\,\rangle$.
Furthermore, Eq.(\ref{timeline-spectral link}) implies that the first term on
the right can be written $\langle\,\tau_{0}\,|\,\psi_{E}\,\rangle\langle
\,\tau\,|\,E\,\rangle/\langle\,\tau_{0}\,|\,E\,\rangle$. Combining these
results, we arrive at an alternative formulation of the eigenvalue problem
posed by Eq.(\ref{eigenvalue problem}) (with $\tau_{0}$ now replaced simply by
$\tau$):%
\begin{equation}
|\,\psi_{E}\,\rangle=\frac{\langle\,\tau\,|\,\psi_{E}\,\rangle}{\langle
\,\tau\,|\,E\,\rangle}\,|\,E\,\rangle+\widehat{G}_{E}\left(  \tau\right)
\widehat{V}|\,\psi_{E}\,\rangle\label{Hilbert}%
\end{equation}
Eq.(\ref{Hilbert}) is more fundamental than its expression in a particular
basis, Eq.(\ref{integral equation}), and will be used exclusively in the
remainder of this paper. We note that the non-uniqueness inherent in the
original Eq.(\ref{eigenvalue problem}) is manifested in Eq.(\ref{Hilbert}) by
the parameter $\tau$ (formerly $\tau_{0}$), which can take any real value.

The inhomogeneous structure of Eq.(\ref{Hilbert}) allows formal solutions to
be found by repeated iteration. Defining $C_{E}\left(  \tau\right)
\equiv\langle\,\tau\,|\,\psi_{E}\,\rangle/\langle\,\tau\,|\,E\,\rangle$, we
get for any $C_{E}\left(  \tau\right)  \neq0$:%
\begin{align*}
|\,\psi_{E}\,\rangle &  =C_{E}\left(  \tau\right)  \,|\,E\,\rangle
+\widehat{G}_{E}\left(  \tau\right)  \widehat{V}C_{E}\left(  \tau\right)
\,|\,E\,\rangle+\ldots\\
&  =\frac{1}{1-\widehat{G}_{E}\left(  \tau\right)  \widehat{V}}\,C_{E}\left(
\tau\right)  |\,E\,\rangle
\end{align*}
Leaving aside questions of convergence, this development convincingly
demonstrates that any state $|\,\psi_{E}\,\rangle$ constructed in this manner
is connected (in the mathematical sense) to the `bare' eigenstate
$|\,E\,\rangle$, and thus persists even in the absence of the interaction
$\widehat{V}$. By contrast, solutions to the corresponding homogeneous
equation%
\begin{equation}
|\,\psi_{E}\,\rangle=\widehat{G}_{E}\left(  \tau\right)  \widehat{V}%
|\,\psi_{E}\,\rangle\label{Hilbert-2}%
\end{equation}
owe their very existence to $\widehat{V}$, vanishing if the interaction is too
weak, or completely absent. We conclude that Eq.(\ref{Hilbert-2}) describes
the quasibound states, defined loosely as those states generated by
$\widehat{V}$ and connected (again in the mathematical sense) through the
growth of some parameter. For consistency, every solution to
Eq.(\ref{Hilbert-2}) must satisfy $\langle\,\tau\,|\,\psi_{E}\,\rangle=0$,
thereby eliminating the first term in Eq.(\ref{Hilbert}); this amounts to a
kind of `initial condition' within the history of states for $\widehat{H}_{0}%
$. We will verify this self-consistency requirement in the next section.

\section{The Quasibound State Green Operator $\protect\widehat{G}_{E}\left(
\tau\right)  $}

From Eq.(\ref{Green's function}) and the closure rule for timeline states,
Eq.(\ref{closure rule}), we obtain the formal representation%
\begin{align*}
\widehat{G}_{E}\left(  \tau\right)   &  =\iint G_{E}\left(  \tau^{\prime
\prime},\tau^{\prime};\tau\right)  \,|\,\tau^{\prime\prime}\,\rangle
\langle\,\tau^{\prime}\,|\,d\tau^{\prime\prime}d\tau^{\prime}\\
&  =i%
{\displaystyle\int\limits_{-\infty}^{\infty}}
d\tau^{\prime\prime}%
{\displaystyle\int\limits_{\tau^{\prime\prime}}^{\tau}}
d\tau^{\prime}\exp\left[  iE\left(  \tau^{\prime\prime}-\tau^{\prime}\right)
\right]  \,|\,\tau^{\prime\prime}\,\rangle\langle\,\tau^{\prime}\,|
\end{align*}
The behavior of timelines under time translation, Eq.(\ref{basis property}),
can be used to write this as%
\begin{align}
\widehat{G}_{E}\left(  \tau\right)   &  =i%
{\displaystyle\int\limits_{-\infty}^{\infty}}
d\tau^{\prime\prime}|\,\tau^{\prime\prime}\,\rangle\langle\,\tau^{\prime
\prime}\,|%
{\displaystyle\int\limits_{\tau^{\prime\prime}}^{\tau}}
d\tau^{\prime}\exp\left[  i\left(  E-\widehat{H}_{0}\right)  \left(
\tau^{\prime\prime}-\tau^{\prime}\right)  \right] \nonumber\\
&  =%
{\displaystyle\int\limits_{-\infty}^{\infty}}
d\tau^{\prime\prime}|\,\tau^{\prime\prime}\,\rangle\langle\,\tau^{\prime
\prime}\,|\frac{1-\exp\left[  i\left(  E-\widehat{H}_{0}\right)  \left(
\tau^{\prime\prime}-\tau\right)  \right]  }{E-\widehat{H}_{0}}\nonumber\\
&  =\left(  1-\frac{|\,E\,\rangle\langle\,\tau\,|}{\langle\,\tau
\,|\,E\,\rangle}\right)  \frac{1}{E-\widehat{H}_{0}} \label{Green-timeline}%
\end{align}
In arriving at the final form, we have invoked closure of the time states,
Eq.(\ref{closure rule}), and the relation (cf.
Eq.(\ref{timeline-spectral link}))%
\[%
{\displaystyle\int\limits_{-\infty}^{\infty}}
d\tau^{\prime\prime}\,|\,\tau^{\prime\prime}\,\rangle\exp\left[  iE\left(
\tau^{\prime\prime}-\tau\right)  \right]  =%
{\displaystyle\int\limits_{-\infty}^{\infty}}
d\tau^{\prime\prime}\,|\,\tau^{\prime\prime}\,\rangle\frac{\langle
\,\tau^{\prime\prime}\,|\,E\,\rangle}{\langle\,\tau\,|\,E\,\rangle}%
=\frac{|\,E\,\rangle}{\langle\,\tau\,|\,E\,\rangle}%
\]

From Eq.(\ref{Green-timeline}) we see at once that $\langle\,\tau
\,|\widehat{G}_{E}\left(  \tau\right)  =0$, thereby confirming that every
solution to the homogeneous Eq.(\ref{Hilbert-2}) complies with the
self-consistency requirement $\langle\,\tau\,|\,\psi_{E}\,\rangle=0$ demanded
by the more general Eq.(\ref{Hilbert}). The same property implies that any
$\langle\,\tau\,|\,\psi_{E}\,\rangle\neq0$ is a non-essential scale factor
(e.g., a normalization choice) whose value can be set independently of
Eq.(\ref{Hilbert}). Evidently, it is the peculiar structure of
Eqs.(\ref{Hilbert}) and (\ref{Green-timeline}) that preserves the identity of
states `sourced' by $\widehat{V}$ even though they be embedded in a continuum,
and it is illuminating to inquire what about this structure is crucial to our
argument, quite apart from the manipulations leading up to it.

Now any procedure that transforms the eigenvalue problem posed by
Eq.(\ref{eigenvalue problem}) into a form like Eq.(\ref{Hilbert}) requires the
operator inverse for $E-\widehat{H}_{0}$. This inverse (the Green operator for
$\widehat{H}_{0}$) we will denote simply by $\widehat{G}_{E}$; it satisfies%
\begin{equation}
\left(  E-\widehat{H}_{0}\right)  \widehat{G}_{E}=1 \label{Green property}%
\end{equation}
But as is well-known, the inverse is not unique, for we can add to
$\widehat{G}_{E}$ any projector of the form $|\,E\,\rangle\langle\,\Omega\,|$
where $|\,E\,\rangle$ is an eigenstate of $\widehat{H}_{0}$ with energy $E$
and $\langle\,\Omega\,|$ is \textit{completely arbitrary}.
Eq.(\ref{Green-timeline}) has just this structure. So in place of
Eqs.(\ref{Hilbert}) and (\ref{Green-timeline}) we are led to consider the
generalized pair
\begin{subequations}
\label{Hilbert-general-pair}%
\begin{align}
|\,\psi_{E}\,\rangle &  =\frac{\langle\,\Gamma\,|\,\psi_{E}\,\rangle}%
{\langle\,\Gamma\,|\,E\,\rangle}|\,E\,\rangle+\widehat{G}_{E}\widehat{V}%
|\,\psi_{E}\,\rangle\label{Hilbert-general-all}\\
\widehat{G}_{E}  &  =\left(  1-\frac{|\,E\,\rangle\langle\,\Gamma\,|}%
{\langle\,\Gamma\,|\,E\,\rangle}\right)  \frac{1}{E-\widehat{H}_{0}}
\label{Green-general}%
\end{align}
with $|\,\Gamma\,\rangle$ so far an unspecified state.

To sort out those states generated by $\widehat{V}$ requires $\langle
\,\Gamma\,|\widehat{G}_{E}=0$, a condition that is automatically satisfied by
Eq.(\ref{Green-general}) provided only that$\,\ \langle\,\Gamma\,|\,E\,\rangle
$ exists and is non-zero. But to achieve faithful sorting, \textit{the
existence and non-vanishing demands on}$\ \langle\,\Gamma\,|\,E\,\rangle
$\textit{ must be met for every energy }$E$\textit{ in the spectrum of
}$\widehat{H}_{0}$\textit{.} For lack of a better term, we will call states
with this property \textit{taggers}; they uniquely identify the eigenstates of
$\widehat{H}_{0}+\widehat{V}$ that are `sourced' by $\widehat{V}$ as having
the property $\langle\,\Gamma\,|\,\psi_{E}\,\rangle=0$.
Eq.(\ref{timeline-spectral link}) shows that time states $|\,\tau\,\rangle
$\ belonging to the history of $\widehat{H}_{0}$ are such taggers. Other
taggers are derivable from these by a restricted class of unitary
transformations: we readily confirm that $|\,\Gamma_{\tau}\,\rangle
=\exp\left[  i\theta(\widehat{H}_{0})\right]  |\,\tau\,\rangle$ with
$\theta(\widehat{H}_{0})$ any real function of $\widehat{H}_{0}$ is also a
tagger, with $\langle\,\Gamma_{\tau}\,|\,E\,\rangle\propto\exp\left(
iE\tau-i\theta\left(  E\right)  \right)  $. Interestingly $|\,\Gamma_{\tau
}\,\rangle$, too, is a time state of $\widehat{H}_{0}$; it differs from
$|\,\tau\,\rangle$ only according to how we fix the phases of the stationary
states $|\,E\,\rangle$. [We have stumbled here upon an important observation:
the quantum history for $\widehat{H}_{0}$\ is a gauge-dependent construct,
with the time states in one gauge related to those in another by a unitary
transformation (more on this later).] That said, we conclude that
\textit{Eqs.(\ref{Hilbert-general-pair}) successfully generalize our previous
results to arbitrary spectra, provided }$|\,\Gamma\,\rangle$ \textit{is
identified with an element in the timeline of }$\widehat{H}_{0}$. The
ambiguity of these time states under gauge transformations notwithstanding, we
simply label them and the associated Green operator with the system time,
writing $|\,\Gamma\,\rangle\rightarrow|\,\tau\,\rangle$ and $\widehat{G}%
_{E}\rightarrow\widehat{G}_{E}(\tau)$. Quasibound states $|\,\psi_{E}%
\,\rangle$ satisfy $\langle\,\tau\,|\,\psi_{E}\,\rangle=0$; equivalently, they
are non-trivial solutions to the homogeneous Eq.(\ref{Hilbert-2}). We note
that degenerate spectra give rise to various kinds of quasibound states,
corresponding to the multiplicity of states $|\,\tau\,\rangle$ for the same
value of $\tau$.

Ubiquitous among the quasibound states referenced in the literature are
quantum resonances, i.e., decaying states with complex energies. If the theory
of quasibound states presented here is to include such resonances, it is not
enough to assume that $E$ is confined to the axis of reals; at a minimum,
$\widehat{G}_{E}\left(  \tau\right)  $ must be extended into those regions of
the complex $E$-plane encompassing the solution space of Eq.(\ref{Hilbert-2}).
To that end, we recognize in $\widehat{G}_{E}\left(  \tau\right)  $ the
\textit{resolvent operator} for the `bare' Hamiltonian $\widehat{H}_{0}$
\cite{Messiah}:%
\end{subequations}
\begin{equation}
\widehat{R}_{0}(E)\equiv\frac{1}{E-\widehat{H}_{0}} \label{resolvent0}%
\end{equation}
The resolvent plays a central role in the formal treatment of quantum
collisions, and its analytical properties as a function of the complex
variable $E$ have been studied extensively. More to the point, the resolvent
operator is intimately tied to scattering resonances: the latter correspond to
poles of the scattering matrix ($S$-matrix), analytically continued into the
complex wavenumber (or energy) plane \cite{Davydov}. These, in turn, coincide
with the poles of the resolvent operator for the full Hamiltonian,
$\widehat{R}(E)\equiv\left(  E-\widehat{H}_{0}-\widehat{V}\right)  ^{-1}$. But
$\widehat{R}(E)$ is related to $\widehat{R}_{0}\left(  E\right)  $ by%
\[
\left(  1-\widehat{R}_{0}\left(  E\right)  \widehat{V}\right)  \,\widehat{R}%
(E)=\widehat{R}_{0}\left(  E\right)  ,
\]
from which we deduce that the \textit{singularities} of $\widehat{R}(E)$ (the
resonance energies) coincide with the \textit{zeros} of $1-\widehat{R}%
_{0}\left(  E\right)  \widehat{V}$. Locating those zeros returns us to an
equation like Eq.(\ref{Hilbert-2}), with $\widehat{G}_{E}\left(  \tau\right)
$ there replaced by $\widehat{R}_{0}\left(  E\right)  $. Thus we are led to
speculate whether $\widehat{G}_{E}\left(  \tau\right)  $ is equivalent to
$\widehat{R}_{0}\left(  E\right)  $ for some choice(s) of $\tau$.

Inspection of Eq.(\ref{Green-timeline}) suggests that $\widehat{G}_{E}\left(
\tau\right)  $ becomes indistinguishable from $\widehat{R}_{0}(E)$ in the
limit of large $\tau$ \textit{provided} $\langle\,\tau\,|\,E\,\rangle$
diverges in this limit. This heuristic argument can be made more precise by
exploiting closure in the spectral basis of $\widehat{H}_{0}$ to write
timeline matrix elements of Eq.(\ref{Green-timeline}) as
\begin{align*}
\langle\,\tau^{\prime}\,|\widehat{R}_{0}(E)-\widehat{G}_{E}\left(
\tau\right)  |\,\tau^{\prime\prime}\,\rangle &  =\frac{\langle\,\tau^{\prime
}\,|\,E\,\rangle}{\langle\,\tau\,|\,E\,\rangle}\langle\,\tau\,|\widehat{R}%
_{0}(E)|\,\tau^{\prime\prime}\,\rangle\\
&  =\frac{\langle\,\tau^{\prime}\,|\,E\,\rangle}{\langle\,\tau\,|\,E\,\rangle}%
{\displaystyle\int}
\frac{dE^{\prime}}{E-E^{\prime}}\langle\,\tau\,|\,E^{\prime}\,\rangle
\,\langle\,E^{\prime}\,|\,\tau^{\prime\prime}\,\rangle\\
&  \propto\exp\left(  iE\tau^{\prime}-iE\tau\right)
{\displaystyle\int}
\frac{\exp\left(  iE^{\prime}\tau-iE^{\prime}\tau^{\prime\prime}\right)
}{E-E^{\prime}}\,dE^{\prime}%
\end{align*}
(Eq.(\ref{timeline-spectral link}) has been invoked in writing the final
form.) So long as $E$ is not real, the integrand in the last line is not
singular anywhere on the path of integration (which extends over all [real]
energies $E^{\prime}$ in the continuous spectrum of $\widehat{H}_{0}$).
Indeed, for $\tau^{\prime\prime}\neq\tau$ this integral converges to a
function that is analytic over the complex $E$-plane, except possibly on the
real axis. Furthermore, due to the rapid fluctuations of the exponential
factor in the integrand, the integral vanishes as $\tau\rightarrow\pm\infty$
for all fixed values of $\tau^{\prime}$ and $\tau^{\prime\prime}$. Over large
swaths of the complex $E$-plane, then, we can expect
\begin{align*}
\lim_{\tau\rightarrow+\infty}\widehat{G}_{E}\left(  \tau\right)   &
=\widehat{R}_{0}\left(  E\right)  \text{\qquad}\Im m\,E<0\\
\lim_{\tau\rightarrow-\infty}\widehat{G}_{E}\left(  \tau\right)   &
=\widehat{R}_{0}\left(  E\right)  \text{\qquad}\Im m\,E>0
\end{align*}
Accordingly, we assert that \textit{the quasibound states defined here for
}$\tau\rightarrow\pm\infty$\textit{ and the scattering resonances encountered
in quantum collision theory are one and the same}.

\section{Quasibound State Selection and Gauge Considerations}

The property of timelines that they span the Hilbert space suggests that
Eqs.(\ref{Hilbert}) and (\ref{Green-timeline}) together furnish an exhaustive
description of all those states generated by the interaction. In particular,
$\psi_{E}(\tau)\equiv\langle\,\tau\,|\,\psi_{E}\,\rangle=0$ for any eigenstate
$|\,\psi_{E}\,\rangle$ of $\widehat{H}_{0}+\widehat{V}$ that owes its
existence to the presence of $\widehat{V}$.

There are several strategies we might employ to calculate quasibound states.
One is to solve the homogeneous relation Eq.(\ref{Hilbert-2}) for some fixed
$\tau$, noting that each such solution automatically satisfies $\psi_{E}%
(\tau)=0$. Another is to solve the eigenvalue problem for the interacting
system, Eq.(\ref{eigenvalue problem}), and supplement the results by imposing
the quasibound state condition $\psi_{E}(\tau)=0$ on the resulting
eigenfunction. Either way, one lingering question remains, viz., how to choose
the value for $\tau$ or, for that matter, how to resolve the phase ambiguity
that breeds distinct (though related) quantum histories? The two are really
the same question, since a changed $\tau$ is itself the result of a gauge
transformation (with $\theta(\widehat{H}_{0})\equiv\Delta\tau\cdot
\widehat{H}_{0}$). And yet, we know a unitary transformation cannot change the
properties of a physical system (in this case, quasibound state energies). So
how do we reconcile this state of affairs? Put succinctly, the quasibound
selection rule in the form $\langle\,\tau\,|\,\psi_{E}\,\rangle=0$ is
\textit{not} manifestly gauge-invariant, which begs the question: can we find
an alternate formulation that is? Further insight into this issue rests on two
additional observations:

\begin{enumerate}
\item Regardless of the specific application, the spectral representation of
time states encapsulated in Eq.(\ref{timeline-spectral link}) together with
closure in the spectral basis generated by $\widehat{H}_{0}$ allows the
quasibound state selection rule to be recast as%
\begin{equation}
\langle\,\tau\,|\,\psi_{E}\,\rangle\propto%
{\displaystyle\int}
\exp\left(  iE^{\prime}\tau\right)  \,\langle\,E^{\prime}\,|\,\psi
_{E}\,\rangle\,dE^{\prime}=0 \label{criterion-general}%
\end{equation}

The integral in Eq.(\ref{criterion-general}) includes all energies $E^{\prime
}$ in the [continuous] spectrum of $\widehat{H}_{0}$, and $\langle\,E^{\prime
}\,|\,\psi_{E}\,\rangle$ is the quasibound wavefunction in the spectral basis
generated by $\widehat{H}_{0}$.

\item For quasibound states with \textit{real energies},
Eq.(\ref{criterion-general}) must admit solutions for real values of $E$, and
this demand constrains the acceptable values we can choose for $\tau$ in that
equation. Henceforth, we will refer to such states as \textit{stationary
quasibound states}, to distinguish them from the resonances so often
identified with quasibound states. We note that stationary quasibound states
are described by physically acceptable wave functions that can be used with
confidence to calculate \textit{all} properties of physical interest.
Precisely this feature was exploited successfully in Ref.\cite{Moyer} to
elicit the divergent behavior of the electric dipole moment accompanying the
onset of electrical breakdown for a bound charge subjected to a uniform
electric field.
\end{enumerate}

To explore the effect of a gauge transformation, let us stipulate first that a
gauge exists in which $\,\langle\,E^{\prime}\,|\,\psi_{E}\,\rangle$ is itself
real-valued for real $E$ and all pertinent [real] values of $E^{\prime}$ (this
appears always to be so, as will become evident from the applications
discussed in subsequent sections). Then Eq.(\ref{criterion-general}) admits
real roots $E$ only if $\tau=0$ (for $\tau\neq0$, complex conjugation results
in a distinct equation that, along with the original, over-constrains the
problem). Equivalently, the selection rule for stationary quasibound states in
this gauge reduces to $\langle\,\tau\,|\,\psi_{E}\,\rangle_{\tau=0}=0$. Next,
consider a gauge transformation that results in the replacement $\,\langle
\,E^{\prime}\,|\,\psi_{E}\,\rangle\rightarrow\exp\left[  i\theta\left(
E^{\prime}\right)  \right]  \,\langle\,E^{\prime}\,|\,\psi_{E}\,\rangle$. To
recover real roots in this instance, we must employ a time state that is a
unitary transform of $|\,0\,\rangle$, namely $\exp\left[  i\theta
(\widehat{H}_{0})\right]  |\,0\,\rangle$. But this leads to the same equation
for $E$ as before (viz., Eq.(\ref{criterion-general}) with $\tau=0$). Thus,
\textit{the specification of stationary quasibound states is indeed
gauge-invariant}, but all conceivable time states must be entertained to
accomodate a general phase assignment when constructing the eigenstates of
$\widehat{H}_{0}$\textit{. }Preserving this flexibility, it would appear, is
the principal role of the parameter $\tau$ in Eq.(\ref{Hilbert}). Conversely,
while complex roots $E$ to Eq.(\ref{criterion-general}) typically exist for
any finite value of $\tau$, they most assuredly \textit{would} be affected by
a phase reassignment (unitary transformation), and therefore cannot represent
any observable property.

The arguments of the preceding paragraph indicate that only in the limits
$\tau\rightarrow\pm\infty$ might we find a gauge-invariant rule for
calculating quasibound states with complex energies. That such states do
indeed exist and are to be identified with the scattering resonances of
quantum collision theory was argued in the preceding section. These, then, are
the only physically permissable outcomes: \textit{stationary quasibound states
may exist, and do so alongside scattering resonances as the only possibilities
for states truly deserving of the branding `quasibound'.}

In the next three sections we will take up quasibound state formation as it
relates to a several distinct choices for $\widehat{H}_{0}$, each specifying
what we call an \textit{application class}. In every instance we exhibit the
time states $|\,\tau\,\rangle$ for that class, and establish the appropriate
[gauge-invariant] selection rule(s) for stationary quasibound states in such
applications. This is followed by calculations for a model potential that
serve to illustrate the formalism in a specific context.

\section{Application Class: The Uniform Field Continuum}

In this -- arguably the simplest -- case, we take $\widehat{H}_{0}%
=\widehat{p}^{2}/2m-F\widehat{x}$, with $F$ denoting the classical force. This
force has the same magnitude and direction everywhere, rendering the problem
essentially one-dimensional. The spectrum is non-degenerate, and stretches
continuously from $E_{\min}=-\infty$ to $E_{\max}=\infty$. While the unbounded
nature of the spectrum from below is considered unphysical, this model
nonetheless serves a useful purpose by sidestepping the issue of boundary
conditions at the potential energy minimum.

Time states for continuous, unbounded spectra are constructed as an ordinary
Fourier transform of the spectral basis states (cf.
Eq.(\ref{timeline-spectral link})). In the coordinate basis,
\begin{equation}
\langle\,x\,|\,\tau\,\rangle\equiv\Xi_{\tau}(x)=\frac{1}{\sqrt{2\pi}}%
{\displaystyle\int\limits_{-\infty}^{\infty}}
\exp\left(  -iE\tau\right)  \,\langle\,x\,|\,E\,\rangle\,dE
\label{history-open}%
\end{equation}
The stationary state waves $\langle\,x\,|\,E\,\rangle$\ for this case are Airy
functions \cite{Landau}; in turn, these give rise to timeline components that
are simply plane waves \cite{Moyer4}:%
\begin{equation}
\Xi_{\tau}(x)=\sqrt{\frac{\left\vert F\right\vert }{2\pi}}\exp\left(
iFx\tau-i\hbar^{2}F^{2}\tau^{3}/6m\right)  \label{waves-open}%
\end{equation}
It is a straightforward matter to confirm that the timeline waves of
Eq.(\ref{waves-open}) are orthogonal, and obey the closure rule of
Eq.(\ref{closure rule}).

\subsection{Stationary Quasibound States in the Uniform Field Continuum}

Quasibound states $|\,\psi_{E}\,\rangle$ are most simply characterized by
$\langle\,\tau\,|\,\psi_{E}\,\rangle=0$. Expressing this condition in the
coordinate basis, we get%
\[%
{\displaystyle\int\limits_{-\infty}^{\infty}}
\,dx\,\Xi_{\tau}(x)\psi_{E}^{\ast}(x)=0
\]
where $\psi_{E}(x)\equiv\langle\,x\,|\,\psi_{E}\,\rangle$ is the
Schr\"{o}dinger waveform for the stationary state of $\widehat{H}%
_{0}+\widehat{V}$ with energy $E$. Apart from an overall scale factor, these
are real functions so long as $\widehat{V}$ has compact support. Thus, the
quasibound selection rule admits real roots $E$ only if $\Xi_{\tau}(x)$ also
is real (apart from a scale factor), which in this gauge requires $\tau=0$, or
(cf. Eq.(\ref{waves-open}))%
\begin{equation}%
{\displaystyle\int\limits_{-\infty}^{\infty}}
\,dx\,\psi_{E}(x)=0 \label{criterion-open}%
\end{equation}
The discussion of Sec. V implies that Eq.(\ref{criterion-open}) is the
gauge-invariant form of the selection rule for stationary quasibound states of
this application class.

The coordinate space version of the quasibound state Green's function
$G_{E}(x,x^{\prime})$ can be found in the usual way as the solution to the
Schr\"{o}dinger equation with a delta function inhomogeneity (cf.
Eq.(\ref{Green property})):%
\[
-\frac{1}{2m}\frac{\partial^{2}G_{E}}{\partial x^{2}}-FxG_{E}(x,x^{\prime
})-EG_{E}(x,x^{\prime})=-\delta(x-x^{\prime})
\]
For $x\neq x^{\prime}$ the solutions are the Airy functions $Ai(-z)$, $Bi(-z)$
of argument $-z$, where $z(x)=\kappa\left(  x+E/F\right)  $ and $\kappa
=\left(  2mF\right)  ^{1/3}$. To complete the specification of $G_{E}%
(x,x^{\prime})$, we impose the selection rule, Eq.(\ref{criterion-open})%
\[%
{\displaystyle\int\limits_{-\infty}^{\infty}}
\,dx\,G_{E}(x,x^{\prime})=0\qquad\forall x^{\prime},
\]
in this way obtaining the uniform-field Green's function for calculating
stationary quasibound states:%
\begin{equation}
G_{E}(x,x^{\prime})=-\frac{\pi\kappa^{2}}{F}Hi(-z^{\prime})Ai(-z)+\frac
{\pi\kappa^{2}}{F}\theta(x-x^{\prime})\left[  Ai(-z)Bi(-z^{\prime
})-Bi(-z)Ai(-z^{\prime})\right]  \label{Green's function-open}%
\end{equation}
Here $z^{\prime}\equiv z(x^{\prime})$, $\theta(\ldots)$ denotes the Heaviside
step function, and $Hi(\ldots)$ is one of two so-called \textit{Scorer}
functions \cite{Scorer}. [The Scorer functions $Gi(z)$ and $Hi(z)$\ are
particular solutions of the inhomogeneous Airy differential equation, and are
related by $Gi(z)+Hi(z)=Bi(z)$.] Details of the calculation leading to
Eq.(\ref{Green's function-open}) have been omitted, since the same result was
reported in an earlier publication \cite{Moyer} using an approach based on the
resolvent operator of Eq.(\ref{resolvent0}) (with $\Im m\,E>0$). For uniform
fields, this resolvent is analytic throughout the whole plane of the complex
variable $E$ \cite{Moyer2}, a property shared by the Green's function of
Eq.(\ref{Green's function-open}).

\subsection{Example: Particle Bound to a Delta Well in a Uniform Field}

To illustrate the computation of quasibound states in the present context, we
consider the delta-function well described by $V(x)=-\lambda\delta(x)$, where
$\lambda>0$ is a measure of the well strength. The delta well supports just
one bound state, with energy $E_{b}=-m\lambda^{2}/2$.

The stationary quasibound states introduced by $V(x)$ satisfy%
\[
\psi_{E}(x)=%
{\displaystyle\int\limits_{-\infty}^{\infty}}
\,dx^{\prime}\,G_{E}(x,x^{\prime})V(x^{\prime})\psi_{E}(x^{\prime})=-\lambda
G_{E}(x,0)\psi_{E}(0)
\]
At $x=0$ this leads to an implicit equation for the energy $E$ of the
quasibound state(s):%
\begin{equation}
1=-\lambda G_{E}(0,0)=2\pi\sqrt{\frac{-\kappa E_{b}}{F}}Hi\left(
-\frac{\kappa E}{F}\right)  Ai\left(  -\frac{\kappa E}{F}\right)
\label{criterion-delta-well}%
\end{equation}
Eq.(\ref{criterion-delta-well}) equation admits exactly one solution with
$E<0$; this root not only coincides with $E_{b}$ in the zero-force limit
$F\rightarrow0$, but actually agrees with the prediction of
Rayleigh--Schr\"{o}dinger perturbation theory \textit{to all orders} in the
force parameter \cite{Moyer3}. While perhaps remarkable on its face, this
agreement should come as no surprise given the essential characteristic of
quasibound states as defined here, viz., their connection to true bound states
through the growth of a parameter (in this case $F$). But the series
development in successively higher powers of $F$ is unmistakenly asymptotic
(and divergent), and so ultimately fails for forces of sufficient strength.
Analysis using (the non-perturbative) Eq.(\ref{criterion-delta-well}) reveals
the existence of a critical strength that results in the destruction of this
quasibound state; the critical value occurs when $E$ reaches zero, and is
given explicitly by \cite{Moyer}%
\begin{equation}
F_{cr}\equiv\left(  2\pi Hi(0)Ai(0)\right)  ^{3}\sqrt{2m}\left(
-E_{b}\right)  ^{3/2}%
\end{equation}
Figs. 1 and 2 show the stationary quasibound state waveform that results for a
force that is 10\% and 90\% of the critical value. For reference, the bound
state wave of the delta well is shown as the dashed curve. All computations
were carried out in units for which $\hbar^{2}=2m=1$. In the weak-force case
(Fig. 1), the stationary quasibound waveform is nearly indistinguishable from
that of the true bound state; by contrast, little semblance of any
localization remains as the critical value is neared (Fig. 2).%
\begin{figure}[ptb]%
\centering
\fbox{\includegraphics[
natheight=2.500200in,
natwidth=3.385700in,
height=2.5278in,
width=3.4134in
]%
{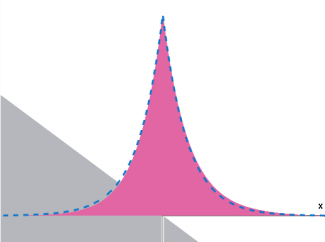}%
}\caption{The stationary quasibound state waveform for a force that is 10\% of
the critical value (the energy of this quasibound state is $E=-1.0955\,E_{b}%
$). The bound state wave (envelope) of the delta well is shown as the dashed
curve, and both are displayed against the backdrop of the full potential
energy curve, $-Fx-\lambda\delta(x)$.}%
\label{figure1}%
\end{figure}
\begin{figure}[ptb]%
\centering
\fbox{\includegraphics[
natheight=2.500200in,
natwidth=3.385700in,
height=2.5278in,
width=3.4134in
]%
{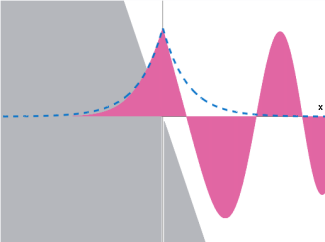}%
}\caption{The stationary quasibound state waveform for a force that is 90\% of
the critical value (the energy of this quasibound state is $E=-0.46404\,E_{b}%
$). The bound state wave (envelope) of the delta well is shown as the dashed
curve, and both are displayed against the backdrop of the full potential
energy curve, $-Fx-\lambda\delta(x)$.}%
\label{figure2}%
\end{figure}

Poles in the complex $E$-plane of the continued resolvent operator for
$\widehat{H}_{0}+\widehat{V}$ (i.e., resonances) also have been studied for
this model potential \cite{Moyer2}, \cite{Ludviksson}-\cite{Alvarez}. Again,
one of these is characterized by a real part that approaches $E_{b}$ as
$F\rightarrow0$ (plus an imaginary part that vanishes exponentially with $F$
in this limit). In fact, the equation for the resonance energies is just
Eq.(\ref{criterion-delta-well}) with $Hi(\ldots)$ $=Bi(\ldots)-Gi(\ldots)$
replaced by the (complex-valued) combination $Bi(\ldots)+iAi(\ldots)$. As
$F\rightarrow0$ with $E<0$, the Airy function arguments tend to infinity,
where $Bi(\ldots)$ dominates over both $Ai(\ldots)$ and $Gi(\ldots)$ to all
orders in the force parameter $F$. It follows that the resonance associated
with the bound state of the delta well and the stationary quasibound state
with $E<0$ cannot be distinguished in the perturbative regime. This feature
actually transcends the example that illustrates it: Ref.\cite{Moyer} implies
that the resonance wave Green's function differs from
Eq.(\ref{Green's function-open}) only by terms that are asymptotically small
compared to any power of $F$, thus assuring identical perturbation series for
any `sourcing' potential $V(x)$. Indeed, the larger point here is one that
underscores the central tenet of our theory: \textit{Stationary quasibound
states and resonances are but different manifestations of those states broadly
termed `quasibound', that derive from true bound states by the growth of a
parameter. The uniqueness of the Rayleigh--Schr\"{o}dinger perturbation series
ensures that both yield identical expansions in that parameter whenever such
developments exist.}

\section{Application Class: The Spherical Wave Continuum}

In this case, the `bare' Hamiltonian $\widehat{H}_{0}$ is the operator for
kinetic energy in a three-dimensional space. The spectrum of $\widehat{H}%
_{0}\ $is semi-infinite (bounded from below by $E=0$, but no upper limit) and
composed of degenerate levels. There is some flexibility in labeling here
depending upon what dynamical variables we opt to conserve along with particle
energy. In the angular momentum representation, the stationary states are
indexed by a continuous wave number $k$ (any non-negative value), an orbital
quantum number $l$ (a non-negative integer), and a magnetic quantum number
$m_{l}$ (an integer between $-l$ and $+l$, not to be confused with particle
mass): $|\,E\,\rangle\rightarrow|\,klm_{l}\,\rangle$. This stationary state
has energy $E_{k}=k^{2}/2m$, and all levels are degenerate with respect to the
angular momentum labels $l$ and $m_{l}$. This application class illustrates
the role of degeneracy in fixing quasibound states when the interaction
$\widehat{V}$ does not mix degenerate states in the continuum.\qquad

The eigenfunctions of $\widehat{H}_{0}$ (the spherical waves) are
$\langle\,\overrightarrow{r}\,|\,klm_{l}\,\rangle\propto$ $\,j_{l}%
(kr)Y_{l}^{m_{l}}(\Omega)$ (product of a spherical Bessel function $j_{l}$
with a spherical harmonic $Y_{l}^{m_{l}}$). The degeneracy in the spectrum of
$\widehat{H}_{0}$ breeds multiple timelines, and these are conveniently
indexed by the same orbital and magnetic quantum numbers that label the
spectral states. In the coordinate basis,%
\begin{equation}
\langle\,\overrightarrow{r}\,|\,\tau\,\rangle\rightarrow\langle
\,\overrightarrow{r}\,|\,\tau lm_{l}\,\rangle=\frac{1}{\sqrt{2\pi}}%
{\displaystyle\int\limits_{0}^{\infty}}
\exp\left(  -iE_{k}\tau\right)  \,\langle\,\overrightarrow{r}\,|\,klm_{l}%
\,\rangle\,dE_{k} \label{history-spherical}%
\end{equation}
We write the time states in the coordinate basis as $\langle
\,\overrightarrow{r}\,|\,\tau lm_{l}\,\rangle\equiv\Xi_{\tau}^{l}%
(r)Y_{l}^{m_{l}}(\Omega)$ where $\Xi_{\tau}^{l}(r)$, the radial piece of the
timeline wave, is given by \cite{Moyer4}%
\begin{equation}
\Xi_{\tau}^{l}(r)=\sqrt{\frac{4r}{m}}z^{3/2}\exp\left(  ir^{2}z-i\pi
\frac{2\alpha+1}{4}\right)  \,\left[  J_{\alpha+1}\left(  r^{2}z\right)
-iJ_{\alpha}\left(  r^{2}z\right)  \right]  \text{\qquad}z\equiv\frac{m}%
{4\tau} \label{waves-BesselJ}%
\end{equation}
Here $J_{\alpha}$ denotes the [cylinder] Bessel function of order $\alpha$,
and $2\alpha\equiv l-1/2$. Eq.(\ref{waves-BesselJ}) is valid for any real
$\tau>0$; to obtain results for $\tau<0$, we use%
\begin{equation}
\Xi_{-\tau}^{l}(r)=\left[  \Xi_{\tau}^{l}(r)\right]  ^{\ast},
\label{time-reversal}%
\end{equation}
a relation suggested by Eq.(\ref{history-spherical}) and established
rigorously in Ref.\cite{Moyer4}.

To further explore the properties of $\Xi_{\tau}^{l}(r)$, it is convenient to
introduce new functions $h_{\alpha}^{\pm}$, related to Hankel functions of the
first and second kind $H_{\alpha}^{(1,2)}$ as
\begin{equation}
h_{\alpha}^{\pm}(z)\equiv\sqrt{\frac{\pi z}{2}}\exp\left(  \mp iz\pm i\pi
\frac{2\alpha+1}{4}\right)  \left[  H_{\alpha+1}^{(1,2)}(z)-iH_{\alpha
}^{(1,2)}(z)\right]  \label{waves-aux}%
\end{equation}
To be clear, $h_{\alpha}^{+}$ involves only Hankel functions of the first
kind, $H^{(1)}$, and $h_{\alpha}^{-}$ only Hankel functions of the second
kind, $H^{(2)}$. Like the Hankel functions, $h_{\alpha}^{\pm}(z)$ are regular
functions of $z$ throughout the $z$-plane cut along the negative real axis.
Using these definitions along with $2J_{\alpha}(z)=H_{\alpha}^{(1)}%
(z)+H_{\alpha}^{(2)}(z)$ allows us to write Eq.(\ref{waves-BesselJ}) as%
\begin{equation}
\Xi_{\tau}^{l}(r)=\sqrt{\frac{2}{\pi mr}}\left[  z\exp\left(  i2r^{2}%
z-i\pi\frac{2\alpha+1}{2}\right)  h_{\alpha}^{+}(r^{2}z)+zh_{\alpha}^{-}%
(r^{2}z)\right]  \text{\qquad}z\equiv\frac{m}{4\tau}, \label{waves-BesselH}%
\end{equation}
with $2\alpha\equiv l-1/2$ as before. The virtue of this representation is
that $h_{\alpha}^{\pm}(\ldots)$ typically are bounded functions with bounded
variation over the entire range of their argument. From their definition, we
find that $h_{\alpha}^{\pm}(z)\sim z^{\alpha+1/2}$ for small $\left\vert
z\right\vert $ and thus bounded as $z\rightarrow0$ provided $\alpha\geq-1/2$.
When $\left\vert z\right\vert $ is large, we employ Hankel's asymptotic series
\cite{Abramowitz2} for the Hankel functions in Eq.(\ref{waves-aux}), in this
way obtaining for any $\alpha>-1/2$ and $\left\vert \arg(z)\right\vert <\pi$:%
\begin{align}
h_{\alpha}^{+}(z)  &  =-2i+\frac{\alpha\left(  \alpha+1\right)  +1/4}%
{z}+o\left(  \frac{1}{z}\right) \nonumber\\
h_{\alpha}^{-}(z)  &  =\frac{\alpha+1/2}{z}+o\left(  \frac{1}{z}\right)
\label{h-bounds}%
\end{align}
In particular, $h_{\alpha}^{+}(z)$ saturates as $\left\vert z\right\vert
\rightarrow\infty$ and $h_{\alpha}^{-}(z)$ vanishes in this limit.

\subsection{Stationary Quasibound States in the Spherical Wave Continuum}

We will assume that the interaction potential has spherical symmetry,
$V(\overrightarrow{r})=V(r)$, so that $l$ and $m_{l}$ remain good quantum
numbers for the eigenstates of $\widehat{H}=\widehat{H}_{0}+\widehat{V}$. Then
the stationary states in the presence of the interaction are $\langle
\,\overrightarrow{r}\,|\,\psi_{E}\,\rangle=R_{El}(r)Y_{l}^{m_{l}}(\Omega)$,
and the quasibound selection rule for this class reduces to%
\begin{equation}%
{\displaystyle\int\limits_{0}^{\infty}}
\,dr\,r^{2}\,\Xi_{\tau}^{l}(r)R_{El}^{\ast}(r)=0
\label{criterion-spherical-all}%
\end{equation}
The limiting forms for $\Xi_{\tau}^{l}(r)$, together with the well-known
properties of the radial wave $R_{El}(r)$ ($R_{El}(r)\sim r^{l}$ as
$r\rightarrow0$ and $rR_{El}(r)$ bounded as $r\rightarrow\infty$
\cite{Merzbacher}) ensure that the integral exists. Moreover, the [effectively
non-degenerate] nature of the spectrum again guarantees that $R_{El}(r)$ is
real, apart from an overall scale factor. We conclude that the quasibound
selection rule is well-formed, and admits real roots $E$ only if $\Xi_{\tau
}^{l}(r)$ also is real (apart from a scale factor). Inspection of
Eq.(\ref{waves-BesselJ}) shows that $\Xi_{\tau}^{l}(r)$ is complex-valued for
any non-zero $\tau$, but also that $\tau=0$ is a singular point for these
functions. Thus we are left to examine Eq.(\ref{criterion-spherical-all}) in
the limit as $\tau\rightarrow0$ in the hope of recovering a viable rule for
distinguishing stationary quasibound states in this application.

To that end, we note that the coefficient of $h_{\alpha}^{+}(\ldots)$ in
Eq.(\ref{waves-BesselH}) is a rapidly varying function of $r$ for $\tau$
small; when substituted into Eq.(\ref{criterion-spherical-all}) we get (after
changing variables to $u=mr^{2}/2$) a term proportional to%
\[
\frac{1}{\tau}%
{\displaystyle\int\limits_{0}^{\infty}}
\,du\,u^{\frac{1}{4}}\exp\left(  iu/\tau\right)  h_{\alpha}^{+}(u/2\tau
)R_{El}(r(u))
\]
Since $h_{\alpha}^{+}$ is bounded for all values of its argument, this last
form is in essence a Fourier integral, whose asymptotics have been thoroughly
studied \cite{Erdelyi}. Assuming that $f(u)\equiv R_{El}(r(u))$ and its
derivatives all exist on $0\leq u<\infty$ and vanish as $u\rightarrow\infty$,
repeated integration by parts generates an asymptotic series in successive
powers of $\tau$. The difficulty here -- if there is one -- comes at the lower
limit ($u=0$), where the inverse transformation $r(u)$ is singular and
derivatives of $f(u)$ may be infinite. To circumvent this problem, we assume
the power series expansion of $R_{El}(r)$ about $r=0$ includes only even,
non-negative powers of $r$, and designate $R_{El}(r)$ as $R_{El}^{+}(r)$ to
emphasize this assumption. To exhibit the leading term in the resulting
asymptotic series for $\tau$, we integrate by parts once using%
\[
v_{\tau}(u)\equiv-%
{\displaystyle\int\limits_{u}^{i\infty}}
dt\,t^{\frac{1}{4}}\exp\left(  it/\tau\right)  h_{\alpha}^{+}(t/2\tau)
\]
The path of integration for $v_{\tau}(u)$ lies entirely in the quadrant
$0\leq\arg(t)\leq\pi/2$, so the integral above converges absolutely. Then%
\begin{align*}
\frac{1}{\tau}%
{\displaystyle\int\limits_{0}^{\infty}}
\,du\,u^{\frac{1}{4}}\exp\left(  iu/\tau\right)  h_{\alpha}^{+}(u/2\tau
)R_{El}^{+}(r(u))  &  \sim-\frac{1}{\tau}R_{El}^{+}(0)v_{\tau}(0)\\
&  =2R_{El}^{+}(0)\left(  2\tau\right)  ^{\frac{1}{4}}%
{\displaystyle\int\limits_{0}^{i\infty}}
dz\,z^{\frac{1}{4}}\exp\left(  i2z\right)  h_{\alpha}^{+}(z)
\end{align*}
This term is at least $O(\tau^{1/4})$ (if $R_{El}^{+}(0)\neq0$) and therefore
vanishes in the limit $\tau\rightarrow0$.

The preceding argument leaves open the question of how to handle any odd
powers in the series expansion of $R_{El}(r)$ about $r=0$. But all odd powers
can be grouped together as $rR_{El}^{+}(r)$, where $R_{El}^{+}(r)$ includes
only even powers of $r$ as before. The prefactor $r$ translates into an extra
factor of $u^{1/2}$, which is absorbed by redefining $v_{\tau}(u)$ as
\[
v_{\tau}(u)\equiv-%
{\displaystyle\int\limits_{u}^{i\infty}}
dt\,t^{\frac{3}{4}}\exp\left(  it/\tau\right)  h_{\alpha}^{+}(t/2\tau)
\]
Then repeated integration by parts again produces an asymptotic series in
$\tau$, now with a leading term at least $O(\tau^{3/4})$. Thus, the main
conclusion reached previously remains intact: such terms make no contribution
to the left side of Eq.(\ref{criterion-spherical-all}) as $\tau\rightarrow0$.

By contrast, the term involving $h_{\alpha}^{-}(\ldots)$ makes a contribution
to Eq.(\ref{criterion-spherical-all}) that is proportional to%
\[%
{\displaystyle\int\limits_{0}^{\infty}}
\,\frac{dr}{\sqrt{r}}\,\,R_{El}(r)r^{2}zh_{\alpha}^{-}(r^{2}z)
\]
Since $r^{2}zh_{\alpha}^{-}(r^{2}z)$ is bounded for all $z$, this integral
converges absolutely and uniformly in $\tau$, so the limit $\tau\rightarrow0$
($z\rightarrow\infty$) may be taken inside the integral to yield the desired
[gauge-invariant] selection rule for stationary quasibound states belonging to
this application class:
\begin{equation}%
{\displaystyle\int\limits_{0}^{\infty}}
\,\frac{dr}{\sqrt{r}}\,R_{El}(r)=0 \label{criterion-spherical}%
\end{equation}

A heuristic argument leading to Eq.(\ref{criterion-spherical}) also can be
given, based on Eq.(\ref{criterion-general}). Using%
\[
\langle\,klm_{l}\,|\,\psi_{E}\,\rangle\propto%
{\displaystyle\int\limits_{0}^{\infty}}
\,\sqrt{k}j_{l}(kr)R_{El}(r)\,r^{2}dr
\]
we substitute into Eq.(\ref{criterion-general}) with $\tau=0$ to get%
\[%
{\displaystyle\int\limits_{0}^{\infty}}
\,\left[
{\displaystyle\int\limits_{0}^{\infty}}
\sqrt{k}j_{l}(kr)\,dE_{k}\right]  R_{El}(r)\,r^{2}dr=0
\]
But%
\[%
{\displaystyle\int\limits_{0}^{\infty}}
\sqrt{k}j_{l}(kr)\,dE_{k}\propto%
{\displaystyle\int\limits_{0}^{\infty}}
j_{l}(kr)k\sqrt{k}\,dk=\frac{1}{r^{2}\sqrt{r}}%
{\displaystyle\int\limits_{0}^{\infty}}
j_{l}(u)u^{3/2}\,du
\]
so Eq.(\ref{criterion-spherical}) is recovered if the integral on the far
right simply exists. Unfortunately $j_{l}$ does not vanish fast enough at
infinity to secure convergence, thus necessitating the more elaborate argument
given previously.

\subsection{Green's Function for Stationary Quasibound Spherical Waves}

The Green's function for this class, $G_{E}^{l}(r,r^{\prime})$, is the
solution to the radial wave equation with a delta function inhomogeneity:%
\[
-\frac{1}{2mr^{2}}\frac{\partial}{\partial r}\left(  r^{2}\frac{\partial
G_{E}^{l}}{\partial r}\right)  +\frac{l\left(  l+1\right)  \hbar^{2}}{2mr^{2}%
}G_{E}^{l}(r,r^{\prime})-EG_{E}^{l}(r,r^{\prime})=-\frac{\delta(r-r^{\prime}%
)}{r^{2}}%
\]
For $r\neq r^{\prime}$ the solutions are spherical Bessel functions
$j_{l}(kr)$, $y_{l}(kr)$, with $k$ related to the particle energy as
$E=k^{2}/2m$. Considered as a function of its first argument, $G_{E}%
^{l}(r,r^{\prime})$ must be regular at the origin, continuous everywhere, but
with a slope discontinuity at $r=r^{\prime}$ as prescribed by the
delta-function singularity there. The solution consistent with these
constraints can be written compactly as%
\begin{equation}
G_{E}^{l}(r,r^{\prime})=\alpha_{l}(r^{\prime})j_{l}(kr)-2mk\,\theta
(r-r^{\prime})\left[  j_{l}(kr)y_{l}(kr^{\prime})-y_{l}(kr)j_{l}(kr^{\prime
})\right]
\end{equation}
where $\theta(\ldots)$\ is the Heaviside step function and $\alpha
_{l}(r^{\prime})$ denotes an as-yet unspecified function. To find $\alpha
_{l}(r^{\prime})$ we impose the selection rule, Eq.(\ref{criterion-spherical}%
); expressed in the language of the Green's function, this is%
\[%
{\displaystyle\int\limits_{0}^{\infty}}
\,\frac{dr}{\sqrt{r}}\,\,G_{E}^{l}(r,r^{\prime})=0\quad\forall r^{\prime}%
\]
From this we obtain the one additional relation that specifies $\alpha
_{l}(r^{\prime})$%
\begin{equation}
\alpha_{l}(r^{\prime})%
{\displaystyle\int\limits_{0}^{\infty}}
\,\frac{du}{\sqrt{u}}\,\,j_{l}(u)=2mk%
{\displaystyle\int\limits_{kr^{\prime}}^{\infty}}
\,\frac{du}{\sqrt{u}}\,\,\left[  j_{l}(u)y_{l}(kr^{\prime})-y_{l}%
(u)j_{l}(kr^{\prime})\right]
\end{equation}
and with it, the complete Green's function for stationary quasibound states in
the spherical wave application class.

In practice, use of the Green's function formulation will require a closed
form for $\alpha_{l}(r^{\prime})$, a task we leave to future investigation.
But the special case of $s$-waves ($l=0$) is sufficiently important and simple
enough to address here. Substituting the spherical Bessel functions of index
zero, and integrating once by parts gives%
\begin{align}
kr^{\prime}%
{\displaystyle\int\limits_{kr^{\prime}}^{\infty}}
\,\frac{du}{\sqrt{u}}\,\left[  j_{0}(u)y_{0}(kr^{\prime})\,-\,y_{0}%
(u)j_{0}(kr^{\prime})\right]   &  =-2%
{\displaystyle\int\limits_{kr^{\prime}}^{\infty}}
\,\frac{du}{\sqrt{u}}\cos\left(  kr^{\prime}-u\right) \nonumber\\
&  =-2\sqrt{2\pi}\,g\left(  \sqrt{\frac{2kr^{\prime}}{\pi}}\right)
\label{fresnelG}%
\end{align}
Here $g(\ldots)$ is the Fresnel auxiliary function \cite{Abramowitz4}, a
non-negative, monotonically decreasing function on $[0,\infty)$ with limit
values $g(0)=1/2$ and $g(\infty)=0$. Also, we can let $kr^{\prime}%
\rightarrow0$ in Eq.(\ref{fresnelG}) to recover the special value%
\[%
{\displaystyle\int\limits_{0}^{\infty}}
\,\frac{du}{\sqrt{u}}\,\,j_{0}(u)=2\sqrt{2\pi}g(0)=\sqrt{2\pi}%
\]
Putting it all together, we obtain the Green's function for stationary
quasibound $s$-waves ($l=0$) in the form%
\begin{equation}
G_{E}^{0}(r,r^{\prime})=-\frac{2m}{r^{\prime}}2g\left(  \sqrt{\frac
{2kr^{\prime}}{\pi}}\right)  \frac{\sin(kr)}{kr}+\frac{2m}{r^{\prime}}%
\,\theta(r-r^{\prime})\frac{\sin(kr-kr^{\prime})}{kr}
\label{Green's function-Swaves}%
\end{equation}

\subsection{Example: $s$-States in a Leaky Spherical Well}

To illustrate quasibound states in the present context, we take for $V(r)$\ a
spherically-symmetric barrier of height $V_{0}>0$ extending from $r=a$ to
$r=b$, and zero elsewhere. Such a barrier effectively creates a
`leaky'\ spherical well of radius $a$ centered at the coordinate origin. There
are no true bound states for this potential, except in the limit of zero
barrier transparency ($b\rightarrow\infty$ and/or $V_{0}\rightarrow\infty$).

Radial wave functions for the $s$-wave ($l=0$) quasibound states introduced by
$V$ satisfy%
\[
R_{E}(r)=%
{\displaystyle\int\limits_{0}^{\infty}}
\,dr^{\prime}\,r^{\prime2}G_{E}^{0}(r,r^{\prime})V(r^{\prime})R_{E}(r^{\prime
})
\]
For $s$-waves, we prefer a formulation in terms of the effective
one-dimensional wave function $u_{E}(r)\equiv rR_{E}(r)$; for the leaky
spherical well, this becomes%
\begin{align*}
u_{E}(r)  &  =V_{0}%
{\displaystyle\int\limits_{a}^{b}}
\,dr^{\prime}r\,r^{\prime}G_{E}^{0}(r,r^{\prime})u_{E}(r^{\prime})\\
&  =-\frac{2mV_{0}}{k}\sin(kr)%
{\displaystyle\int\limits_{a}^{b}}
\,dr^{\prime}2g\left(  z(r^{\prime})\right)  u_{E}(r^{\prime})+\frac{2mV_{0}%
}{k}%
{\displaystyle\int\limits_{a}^{b}}
\,dr^{\prime}\theta(r-r^{\prime})\sin(kr-kr^{\prime})u_{E}(r^{\prime})
\end{align*}
where for brevity we have introduced $z(r)\equiv\sqrt{2kr/\pi}$. We see
immediately that $u_{E}(r)\propto\sin(kr)$ for $r\leq a$, as expected. For
$a\leq r\leq b$ and $V_{0}>E$, $u_{E}(r)$ should be a combination of growing
and decaying exponentials $\exp\left(  \pm\kappa r\right)  $ with $\kappa
^{2}=2m\left(  V_{0}-E\right)  $. If, in the barrier region, we write%
\[
u_{E}(r)=\exp(-\kappa r)+s\exp(\kappa r)\qquad a\leq r\leq b
\]
then continuity of the logarithmic derivative of $u_{E}(r)$ at $r=a$ specifies
the mixing coefficient $s$ as%
\begin{equation}
s=\exp(-2\kappa a)\frac{\kappa\sin(ka)+k\cos(ka)}{\kappa\sin(ka)-k\cos(ka)}
\label{mixing coefficient}%
\end{equation}
Enforcing continuity of $u_{E}(r)$ at $r=a$ then leads to%
\[
\exp(-\kappa a)+s\exp(\kappa a)=-\frac{2mV_{0}}{k}\sin(ka)%
{\displaystyle\int\limits_{a}^{b}}
\,dr\,2g\left(  z(r)\right)  \left[  \exp(-\kappa r)+s\exp(\kappa r)\right]
\]
Finally, substituting for $s$ from Eq.(\ref{mixing coefficient}) and
rearranging gives%
\begin{align}
-1  &  =\left[  \frac{\kappa}{k}\sin(ka)-\cos(ka)\right]  \,\frac{2mV_{0}%
}{\kappa}%
{\displaystyle\int\limits_{a}^{b}}
\,dr\,g\left(  z(r)\right)  \exp(\kappa a-\kappa r)\nonumber\\
&  +\left[  \frac{\kappa}{k}\sin(ka)+\cos(ka)\right]  \,\frac{2mV_{0}}{\kappa}%
{\displaystyle\int\limits_{a}^{b}}
\,dr\,g\left(  z(r)\right)  \exp(\kappa r-\kappa a)
\label{criterion-leaky-well}%
\end{align}
Despite appearances, $k$ and $\kappa$ are not independent here, but linked for
a given potential barrier by the familiar relation
\[
k^{2}+\kappa^{2}=2mV_{0}%
\]

Eq.(\ref{criterion-leaky-well}) as written is valid for $E<V_{0}$; it is an
implicit equation whose roots $ka$ prescribe the stationary quasibound states
with below-the-barrier energy in this leaky well. Roots must be found
numerically, but several noteworthy observations can be made without further computation:

\begin{enumerate}
\item In the thick barrier limit ($b\rightarrow\infty$) the last integral on
the right of Eq.(\ref{criterion-leaky-well}) diverges, whereas the remaining
terms are bounded for all $k,\kappa>0$. It follows that the coefficient of the
divergent integral must vanish in this limit. Eq.(\ref{criterion-leaky-well})
further suggests that this coefficient approaches zero exponentially in $b$.
Indeed, careful analysis shows that for large $b$ ($w\equiv b-a$ is the
barrier width)%
\[
\frac{\kappa}{k}\sin(ka)+\cos(ka)\sim b^{3/2}\exp\left(  -\kappa w\right)
\]
Since the zeros of the expression on the left specify the bound state energies
of a spherical well with width $a$, we conclude that the stationary quasibound
levels merge with those bound state levels as $b\rightarrow\infty$. Also since
there is no degeneracy in the $s$-wave spectrum -- either with or without the
`sourcing' potential $V(r)$ -- it follows that the stationary quasibound
waveforms themselves converge to the bound state wave functions of the
spherical well in this limit.

\item As the barrier narrows, the right side of Eq.(\ref{criterion-leaky-well}%
) steadily shrinks in magnitude until that equation is incapable of supporting
solutions. To explore this point further, we rewite
Eq.(\ref{criterion-leaky-well}) in the abbreviated form%
\[
-1=A\frac{\sin(ka)}{ka}+B\cos(ka)
\]
with obvious definitions for $A$ and $B$. Since $g\left(  z\right)  $ is a
monotonically decreasing function, we easily deduce the bounds
\begin{align}
A  &  \lesssim2mV_{0}aw\frac{\sinh\left(  \kappa w\right)  }{\left(  \kappa
w\right)  }2g\left(  z_{a}\right)  =2mV_{0}aw\frac{\sinh\left(  \kappa
w/2\right)  \cosh\left(  \kappa w/2\right)  }{\left(  \kappa w/2\right)
}2g\left(  z_{a}\right) \nonumber\\
B  &  \lesssim2mV_{0}w^{2}\frac{\cosh\left(  \kappa w\right)  -1}{\left(
\kappa w\right)  ^{2}}2g\left(  z_{a}\right)  =2mV_{0}w^{2}\left[  \frac
{\sinh\left(  \kappa w/2\right)  }{\left(  \kappa w/2\right)  }\right]
^{2}g\left(  z_{a}\right)  \label{bounds-AB}%
\end{align}
For a thin barrier ($\tanh\left(  \kappa w/2\right)  \ll\kappa a/2$), $A$
appears to be dominant and the equation for the quasibound energies becomes,
approximately,%
\[
-1\simeq2mV_{0}aw\frac{\sinh\left(  \kappa w\right)  }{\left(  \kappa
w\right)  }\left[  2g\left(  \sqrt{\frac{2ka}{\pi}}\right)  \frac{\sin
(ka)}{ka}\right]
\]
Considered as a function of $ka$, the bracketed term on the right is a
steadily decaying oscillation about zero. Numerical investigation shows that
the first (deepest) minimum occurs at $ka\simeq4.2149$, and the function value
at this point is $\approx-1/120$. It follows that no stationary quasibound
states can exist if%
\begin{equation}
2mV_{0}aw\frac{\sinh\left(  \kappa w\right)  }{\left(  \kappa w\right)  }%
\leq\sqrt{2mV_{0}a^{2}}\sinh\left(  \sqrt{2mV_{0}w^{2}}\right)  \lesssim120
\label{cutoff-thin barrier}%
\end{equation}

Although not rigorous, the bound of Eq.(\ref{cutoff-thin barrier}) is
sufficiently accurate as to be quite useful in practice.
\end{enumerate}

The integrals appearing in Eq.(\ref{criterion-leaky-well}) can actually be
done in closed form with the help of the following [indefinite] integral:%
\begin{equation}
\left[  1+\frac{\pi^{2}}{4\gamma^{2}}\right]
{\displaystyle\int^{z}}
\,d(\gamma z^{2})\,g\left(  z\right)  \exp(\gamma z^{2})=\left[  g\left(
z\right)  -\frac{\pi}{2\gamma}f(z)\right]  \exp(\gamma z^{2})+i\sqrt{\frac
{\pi}{4\gamma}}\operatorname{erf}\left(  -i\sqrt{\gamma}\,z\right)
\end{equation}
Here $f(z)$ is the Fresnel auxiliary function companion to $g(z)$
\cite{Abramowitz4}, and $\operatorname{erf}\left(  \ldots\right)  $ denotes
the familiar error function. This result allows direct numerical computation
of the stationary quasibound levels for this example. A thorough analysis is
out of place here, but may appear in a future publication.

Resonances for twin symmetric barriers have been investigated by Maheswari,
et. al. \cite{Maheswari}. The antisymmetric states in their study become the
$s$-wave resonances for the leaky\ well described here. They report numerical
results for the lowest 8-10 resonance energies (along with their widths)
generated by a relatively thin ($w=0.167\,a$), as well as a moderately thick
($w=0.5\,a$) rectangular barrier, with $V_{0}a^{2}=72$ (in units where
$\hbar^{2}=2m=1$) for both. Only the latter is thick enough to support any
stationary quasibound states. [With $V_{0}a^{2}=72$, the critical barrier
thickness implied by Eq.(\ref{cutoff-thin barrier}) is $w_{cr}\simeq0.394\,a$,
whereas the actual value $w_{cr}\simeq0.425\,a$ is about $7\%$ larger.] Our
calculations show that for $V_{0}a^{2}=72$ and $w=0.5\,a$ the lowest-lying
stationary quasibound state has energy $E=1.067$. And there is just one
additional stationary quasibound state for these parameter values, at energy
$E=2.331$. For comparison, Maheswari, et. al. find -- using these same
parameters -- three antisymmetric resonances below the top of the barrier
($E_{r}<V_{0}$), at energies $E_{r}=0.874$, $3.444$, and $7.421$.%
\begin{figure}[ptb]%
\centering
\fbox{\includegraphics[
natheight=2.500200in,
natwidth=3.385700in,
height=2.5278in,
width=3.4134in
]%
{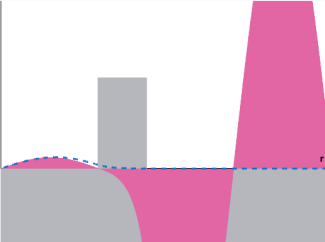}%
}\caption{The lowest-lying quasibound $s$-state for a spherically-symmetric
barrier, with $V_{0}a^{2}=72$ and $w=0.5\,a$. The bound state wave (envelope)
of the spherical well that results from an infinitely thick barrier is shown
as the dashed curve, and both are displayed against the backdrop of the
barrier potential $V(r)$.}%
\label{figure3}%
\end{figure}

The differences between stationary quasibound states and resonances are
evident in these results, perhaps in no small part because a perturbative
treatment is not possible here, and because with the chosen parameters even
the most stable of the stationary quasibound states is on the verge of
extinction. The wave function for the lowest-lying stationary quasibound state
is shown in Fig. 3, where it is compared with the bound state wave to which it
converges in the thick barrier limit $b\rightarrow\infty$. The resemblance is
clear inside the spherical well ($0\leq r\leq a$) and just beyond, but there
the similarities end. The bound state wave continues to decay exponentially
with $r$, while the quasibound waveform emerges at the outer barrier edge with
considerable amplitude and is oscillatory beyond. [The external wave amplitude
actually grows with increasing $b$, even as the waveform up to and including
the first node converges smoothly to the bound state wave.] With these same
parameters, the associated resonance has energy (real part) indistinguishable
from that of the bound state ($E_{r}=0.874$) to the accuracy reported, making
the resonance and bound state wave functions virtually identical over the
range shown. Still, the imaginary part of the resonance energy -- though quite
small ($E_{i}=-\,0.718\times10^{-4}$) -- causes the resonance wave to diverge
exponentially in the asymptotic region $r\rightarrow\infty$.

\section{Application Class: The Free Particle Continuum in One Dimension}

For this case $\widehat{H}_{0}$ describes a particle free to move along the
line $-\infty<x<\infty$. The eigenfunctions of $\widehat{H}_{0\text{ }}$are
harmonic waves with wavenumber $k$ and energy $E_{k}=k^{2}/2m$. Since $k$ can
be any real number, the spectrum extends from $E_{k}=0$ to $E_{k}=\infty$ and
each energy level is doubly-degenerate ($\pm k$). This application class
illustrates how spectral degeneracy affects the search for stationary
quasibound states when the interaction giving rise to those states actually
mixes degenerate states in the continuum.

The eigenfunctions may be taken as plane waves, and plane waves running in
opposite directions give rise to distinct quantum histories. Alternatively,
histories can be constructed from standing wave combinations of these plane
waves, and is the course we follow here. Timeline elements in the standing
wave picture are described by the coordinate-space forms $\langle
\,x\,|\,\tau\pm\rangle\equiv\Xi_{\tau}^{\pm}(x)$ where the sign label $\left(
\pm\right)  $ specifies the parity of these waves.

The odd-parity states are simply related to the timeline states for spherical
waves with $l=0$; indeed, $\sqrt{2}\,\Xi_{\tau}^{-}(x)=x\,\Xi_{\tau}^{0}(x)$
\cite{Moyer4}. Applying the result of Eq.(\ref{waves-BesselH}) to this case
($\alpha=-1/4$), we obtain for $x\geq0$ and $\tau>0$
\begin{equation}
\Xi_{\tau}^{-}(x)=\sqrt{\frac{x}{\pi m}}\left[  z\exp\left(  i2x^{2}%
z-i\pi/4\right)  h_{-1/4}^{+}(x^{2}z)+zh_{-1/4}^{-}(x^{2}z)\right]
\text{\qquad}z\equiv\frac{m}{4\tau} \label{waves-free-odd}%
\end{equation}
It is noteworthy that the odd-parity states by themselves constitute a
complete history for an otherwise free particle that is confined to the
half-axis $x>0$ (by an infinite potential wall at the origin). But for a truly
free particle we also need the even-parity states.

For the even-parity states, we exploit the formal connection to time states
for spherical waves with $l=-1$: $\sqrt{2}\,\Xi_{\tau}^{+}(x)=x\,\Xi_{\tau
}^{-1}(x)$ \cite{Moyer4}. While unphysical for spherical waves,
Eq.(\ref{waves-BesselH}) continues to hold for $l=-1$ ($\alpha=-3/4$) and
gives for $x\geq0$ and $\tau>0$%
\begin{equation}
\Xi_{\tau}^{+}(x)=\sqrt{\frac{x}{\pi m}}\left[  z\exp\left(  i2x^{2}%
z+i\pi/4\right)  h_{-3/4}^{+}(x^{2}z)+zh_{-3/4}^{-}(x^{2}z)\right]
\text{\qquad}z\equiv\frac{m}{4\tau} \label{waves-free-even}%
\end{equation}
The extension of these results to $\tau<0$ is afforded by
Eq.(\ref{time-reversal}); the extension to $x<0$ is dictated by parity.

\subsection{Stationary Quasibound States in the Free-Particle Continuum}

In the present context, there are a pair of enforceable conditions leading to
two distinct kinds of quasibound states: they are%
\begin{equation}%
{\displaystyle\int\limits_{-\infty}^{\infty}}
\,dx\,\,\Xi_{\tau}^{\pm}(x)\psi_{E}^{\ast}(x)=0 \label{criterion-free-all}%
\end{equation}
where again $\psi_{E}(x)\equiv\langle\,x\,|\,\psi_{E}\,\rangle$ is the
Schr\"{o}dinger waveform for the stationary state of $\widehat{H}%
_{0}+\widehat{V}$ with energy $E$. So long as $\psi_{E}(x)$ is everywhere
bounded, the integrals of Eq.(\ref{criterion-free-all}) exist and the
quasibound state criteria are well-formed. While $\psi_{E}(x)$ is not
guaranteed to be real, its complex conjugate is a stationary state wave with
the same energy. Thus, we may form two new states $\psi_{E}(x)\pm\psi
_{E}^{\ast}(x)$ which \textit{are} real up to a constant multiplier. In
effect, without loss of generality we can always assume $\psi_{E}(x)$ is real
(apart from a constant multiplier) in Eq.(\ref{criterion-free-all}) for the
purpose of computing quasibound states. Furthermore, the candidate wave
function can always be split into even and odd parts, $\psi_{E}(x)=\psi
_{E}^{+}(x)+\psi_{E}^{-}(x)$, such that%
\[
\psi_{E}^{\pm}(x)=\frac{1}{2}\left[  \psi_{E}(x)\pm\psi_{E}(-x)\right]
\]
are themselves real (again apart from a constant multiplier). With this
decomposition, the quasibound criteria of Eq.(\ref{criterion-free-all}) reduce
to%
\begin{equation}%
{\displaystyle\int\limits_{0}^{\infty}}
\,dx\,\,\Xi_{\tau}^{\pm}(x)\psi_{E}^{\pm}(x)=0 \label{criterion-free-reduced}%
\end{equation}
where either the upper or lower signs are taken together. The odd-parity case
($-$) is formally identical to that for spherical $s$-waves with the
replacement $R_{El}(r)\rightarrow$ $x^{-1}\psi_{E}^{-}(x)$; thus for
$u=mx^{2}/2$, so long as $f(u)\equiv\left[  x(u)\right]  ^{-1}\psi_{E}%
^{-}(x(u))$ and its derivatives all exist on $0\leq u<\infty$ and vanish as
$u\rightarrow\infty$, we are led in the same way to one [gauge-invariant]
selection rule for stationary quasibound states belonging to this application
class:%
\begin{equation}%
{\displaystyle\int\limits_{0}^{\infty}}
\,\frac{dx}{\sqrt[3]{x}}\,\psi_{E}^{-}(x)=0 \label{criterion-free-odd}%
\end{equation}

The even-parity case is a bit trickier. Let us define%
\[
I_{\tau}(x)\equiv%
{\displaystyle\int\limits_{0}^{x}}
\Xi_{\tau}^{+}(x^{\prime})\,dx^{\prime}%
\]
Using the properties of Bessel functions, it is a straightforward exercise to
show that this integral can be evaluated in terms of $\Xi_{\tau}^{+}(x)$ and
the time state for $p$-waves, $\Xi_{\tau}^{1}(x)$:
\[
I_{\tau}(x)=x\Xi_{\tau}^{+}(x)+\frac{x^{2}}{\sqrt{2}}\Xi_{\tau}^{1}(x)
\]
Then from Eq.(\ref{waves-free-even}) and the spherical wave result of
Eq.(\ref{waves-BesselH}) with $l=1$ ($\alpha=1/4$), we obtain a corresponding
expression for $I_{\tau}(x)$:
\begin{equation}
I_{\tau}(x)=\sqrt{\frac{x^{3}}{\pi m}}\left[  \exp\left(  i2x^{2}%
z+i\pi/4\right)  z\left(  h_{-3/4}^{+}(x^{2}z)-h_{1/4}^{+}(x^{2}z)\right)
+zh_{-3/4}^{-}(x^{2}z)+zh_{1/4}^{-}(x^{2}z)\right]
\label{waves-free_alternate}%
\end{equation}
The asymptotic results for $h^{\pm}$ from Eq.(\ref{h-bounds}) show that
$I_{\tau}(x)$ vanishes as $x\rightarrow\infty$; thus,
Eq.(\ref{criterion-free-reduced}) can be integrated once by parts to express
the quasibound state criterion for the even-parity case in the alternate form%
\begin{equation}%
{\displaystyle\int\limits_{0}^{\infty}}
\,dx\,I_{\tau}(x)\frac{d\psi_{E}^{+}}{dx}=0 \label{criterion-free-even-all}%
\end{equation}

From this point on, the argument parallels that given previously for the case
of spherical waves. Real energy solutions to Eq.(\ref{criterion-free-even-all}%
) demand that $I_{\tau}(x)$ be real and this, in turn, requires $\tau
\rightarrow0$. The coefficient of $h_{\alpha}^{+}(\ldots)$ ($\alpha
=-3/4,\,1/4$) in Eq.(\ref{waves-free_alternate}) is a rapidly-varying function
of $x$ for $\tau$ small; when substituted into
Eq.(\ref{criterion-free-even-all}) these generate terms proportional to (after
changing variables to $u=mx^{2}/2$)%
\[
\frac{1}{\tau}%
{\displaystyle\int\limits_{0}^{\infty}}
\,du\,u^{\frac{3}{4}}\exp\left(  iu/\tau\right)  h_{\alpha}^{+}(u/2\tau
)\left[  \frac{1}{x}\frac{d\psi_{E}^{+}}{dx}\right]
\]
So long as $f(u)\equiv\left[  x(u)\right]  ^{-1}d\psi_{E}^{+}/dx$ and its
derivatives all exist on $0\leq u<\infty$ and vanish as $u\rightarrow\infty$,
repeated integration by parts generates an asymptotic series in successive
powers of $\tau$. To obtain the leading term in that series, we integrate by
parts once using%
\[
v_{\tau}(u)\equiv-%
{\displaystyle\int\limits_{u}^{i\infty}}
dt\,t^{\frac{3}{4}}\exp\left(  it/\tau\right)  h_{\alpha}^{+}(t/2\tau)
\]
with the integration contour confined to the quadrant $0\leq\arg(t)\leq\pi/2$.
Then%
\begin{align*}
\frac{1}{\tau}%
{\displaystyle\int\limits_{0}^{\infty}}
\,du\,u^{\frac{3}{4}}\exp\left(  iu/\tau\right)  h_{\alpha}^{+}(u/2\tau
)\left[  \frac{1}{x}\frac{d\psi_{E}^{+}}{dx}\right]   &  \sim-\frac{1}{\tau
}\left[  \frac{1}{x}\frac{d\psi_{E}^{+}}{dx}\right]  _{0}v_{\tau}(0)\\
&  =\left[  \frac{1}{x}\frac{d\psi_{E}^{+}}{dx}\right]  _{0}2\left(
2\tau\right)  ^{\frac{3}{4}}%
{\displaystyle\int\limits_{0}^{i\infty}}
dz\,z^{\frac{3}{4}}\exp\left(  i2z\right)  h_{\alpha}^{+}(z)
\end{align*}
Evidently such contributions are at least $O\left(  \tau^{3/4}\right)  $, and
therefore vanish in the limit $\tau\rightarrow0$. By contrast, the remaining
terms in Eq.(\ref{waves-free_alternate}) make a contribution proportional to%
\[%
{\displaystyle\int\limits_{0}^{\infty}}
\,dx\,\sqrt{x^{3}}\,\left[  zh_{-3/4}^{-}(x^{2}z)+zh_{1/4}^{-}(x^{2}z)\right]
\frac{d\psi_{E}^{+}}{dx}%
\]
Since $x^{2}zh_{\alpha}^{-}(x^{2}z)$ is bounded for all $z$, this integral
converges uniformly in $\tau$ (but not absolutely), so the limit
$\tau\rightarrow0$ ($z\rightarrow\infty$) may be taken inside the integral to
yield a second [gauge-invariant] selection rule for stationary quasibound
states in this application class:%
\begin{equation}%
{\displaystyle\int\limits_{0}^{\infty}}
\,\frac{dx}{\sqrt{x}}\,\,\frac{d\psi_{E}^{+}}{dx}=0
\label{criterion-free-even}%
\end{equation}

Interestingly, the first selection rule, Eq.(\ref{criterion-free-odd}), can be
cast in identical form with the help of a single integration by parts. Thus,
both quasibound selection rules for this application class can be expressed by
the single compact equation%
\begin{equation}%
{\displaystyle\int\limits_{0}^{\infty}}
\,\frac{dx}{\sqrt{x}}\,\,\frac{d\psi_{E}^{\pm}}{dx}=0
\label{criterion-free-both}%
\end{equation}
where $\psi_{E}^{\pm}(x)$ are the even ($+$) and odd ($-$) parts of the
stationary state wave function with energy $E$.

\subsection{Green's Function Formulation}

The Green's function $G_{E}(x,x^{\prime})$ for this case is the solution to
the free-particle Schr\"{o}dinger equation with a delta function
inhomogeneity. Like the stationary states, $G_{E}(x,x^{\prime})$ can be split
into components that are even and odd under reflection (in the first
argument); we write this as $G_{E}(x,x^{\prime})=G_{E}^{+}(x,x^{\prime}%
)+G_{E}^{-}(x,x^{\prime})$ where%
\[
G_{E}^{\pm}(x,x^{\prime})=\frac{1}{2}\left[  G_{E}(x,x^{\prime})\pm
G_{E}(-x,x^{\prime})\right]
\]
clearly has the property $G_{E}^{\pm}(-x,x^{\prime})=\pm G_{E}^{\pm
}(x,x^{\prime})$. These component Green's functions satisfy the differential
equation%
\begin{equation}
-\frac{1}{2m}\frac{\partial^{2}G_{E}^{\pm}}{\partial x^{2}}-EG_{E}^{\pm
}(x,x^{\prime})=-\frac{1}{2}\left[  \delta(x-x^{\prime})\pm\delta(x+x^{\prime
})\right]  \label{Green's function equation}%
\end{equation}
For stationary quasibound states, solutions also must conform to the criteria
(cf. Eq.(\ref{criterion-free-both}))%
\begin{equation}%
{\displaystyle\int\limits_{0}^{\infty}}
\,\frac{dx}{\sqrt{x}}\frac{\partial G_{E}^{\pm}(x,x^{\prime})}{\partial
x}\,=0\qquad\forall x^{\prime} \label{Green-criteria-both}%
\end{equation}
Changing the sign of $x^{\prime}$ in Eqs.(\ref{Green's function equation}) and
(\ref{Green-criteria-both}) implies that $G_{E}^{\pm}(x,-x^{\prime})$ $=\pm
G_{E}^{\pm}(x,x^{\prime})$, i.e., \textit{the component Green's functions
exhibit the same symmetry under reflection in the second argument as they do
for the first argument}. It follows that we need only construct $G_{E}^{\pm
}(x,x^{\prime})$ over the range $x,x^{\prime}\geq0$.

Consider first the odd-parity component $G_{E}^{-}(x,x^{\prime})$. For
$x,x^{\prime}\geq0\ $this is intimately related to the Green's function for
spherical $s$-waves $G_{E}^{0}(r,r^{\prime})$: indeed, we easily discover that
$2G_{E}^{-}(x,x^{\prime})=$ $xx^{\prime}G_{E}^{0}(x,x^{\prime})$. More
precisely (cf. Eq.(\ref{Green's function-Swaves})),
\begin{equation}
G_{E}^{-}(x,x^{\prime})=-\frac{m}{k}2g\left(  \sqrt{\frac{2kx^{\prime}}{\pi}%
}\right)  \sin(kx)+\frac{m}{k}\,\theta(x-x^{\prime})\sin(kx-kx^{\prime})\qquad
x,x^{\prime}\geq0 \label{Green's function-free-odd}%
\end{equation}

For the even-parity component $G_{E}^{+}(x,x^{\prime})$, solutions to
Eq.(\ref{Green's function equation}) for $x\neq x^{\prime}$ are free-particle
standing waves with wavenumber $k$ and energy $E=k^{2}/2m$ that have vanishing
slope at $x=0$. For $x,x^{\prime}\geq0$ we write this as%
\[
G_{E}^{+}(x,x^{\prime})=%
\begin{array}
[c]{ccc}%
\alpha(x^{\prime})\cos(kx) & \qquad & 0\leq x\leq x^{\prime}\\
\beta(x^{\prime})\cos(kx)+\gamma(x^{\prime})\sin(kx) & \qquad & x>x^{\prime
}\geq0
\end{array}
\]
Continuity at $x=x^{\prime}$ requires%
\[
\left[  \alpha(x^{\prime})-\beta(x^{\prime})\right]  \cos(kx^{\prime}%
)=\gamma(x^{\prime})\sin(kx^{\prime}),
\]
while integrating the differential equation across the singular point at
$x=x^{\prime}$ gives
\[
\left[  \beta(x^{\prime})-\alpha(x^{\prime})\right]  \frac{d}{dx^{\prime}}%
\cos(kx^{\prime})+\gamma(x^{\prime})\frac{d}{dx^{\prime}}\sin(kx^{\prime})=m
\]
Combining these matching conditions with the Wronskian $W\left\{  \cos z,\sin
z\right\}  =1$ leads to%
\begin{align*}
\gamma(x^{\prime})  &  =\frac{m}{k}\cos(kx^{\prime})\\
\alpha(x^{\prime})-\beta(x^{\prime})  &  =\frac{m}{k}\sin(kx^{\prime})
\end{align*}

The specification of $G_{E}^{+}(x,x^{\prime})$\ is completed by imposing the
selection rule for stationary quasibound states, Eq.(\ref{Green-criteria-both}%
). After some manipulation, we find for any value of $x^{\prime}\geq0$%
\[
\alpha(x^{\prime})%
{\displaystyle\int\limits_{0}^{\infty}}
\,\frac{du}{\sqrt{u}}\sin u=\frac{m}{k}%
{\displaystyle\int\limits_{kx^{\prime}}^{\infty}}
\,\frac{du}{\sqrt{u}}\cos\left(  u-kx^{\prime}\right)
\]
The integral on the left is related to the Fresnel sine integral $S(\infty)$
and evaluates to $\sqrt{2\pi}$ \cite{Abramowitz4}, while the one on the right
is essentially the Fresnel auxiliary function $g(\ldots)$ introduced in Sec.
VII (cf. Eq.(\ref{fresnelG})). Collecting all the above results leads to the
Green's function for even-parity stationary quasibound states in the form%
\begin{equation}
G_{E}^{+}(x,x^{\prime})=\frac{m}{k}\,2g\left(  \sqrt{\frac{2kx^{\prime}}{\pi}%
}\right)  \cos(kx)+\frac{m}{k}\,\theta(x-x^{\prime})\sin(kx-kx^{\prime})\qquad
x,x^{\prime}\geq0 \label{Green's function-free-even}%
\end{equation}

\subsection{Example: Twin Rectangular Barriers}

We consider here a pair of identical rectangular barriers located a distance
$a$ to either side of the coordinate origin. The potential energy $V(x)$\ is
constant at $V_{0}>0$ within the barrier regions $-b\leq x\leq-a$ and $a\leq
x\leq b$, and zero elsewhere. The barrier thickness is $w=b-a$. The twin
barriers effectively delineate the edges of a `leaky'\ well of width $2a$
centered at the coordinate origin.

Since $V(x)$ has reflection symmetry about the origin, the stationary states
can be taken even or odd under reflection, and only the half-axis $x\geq0$
need be considered. Stationary quasibound states having odd parity must
satisfy the integral equation%
\[
\psi_{E}^{-}(x)=2%
{\displaystyle\int\limits_{0}^{\infty}}
\,dx^{\prime}\,G_{E}^{-}(x,x^{\prime})V(x^{\prime})\psi_{E}^{-}(x^{\prime})
\]
This is the same problem encountered in the previous section for $s$-wave
quasibound states in a leaky spherical well of radius $a$, and leads to
identical results; accordingly, we will not pursue the odd-parity case any
further here.

For the even-parity case, we have simil;arly%
\begin{align*}
\psi_{E}^{+}(x)  &  =2%
{\displaystyle\int\limits_{0}^{\infty}}
\,dx^{\prime}\,G_{E}^{+}(x,x^{\prime})V(x^{\prime})\psi_{E}^{+}(x^{\prime})\\
&  =\frac{2mV_{0}}{k}\cos(kx)%
{\displaystyle\int\limits_{a}^{b}}
\,dx^{\prime}2g\left(  \sqrt{\frac{2kx^{\prime}}{\pi}}\right)  \psi_{E}%
^{+}(x^{\prime})+\frac{2mV_{0}}{k}%
{\displaystyle\int\limits_{a}^{b}}
\,dx^{\prime}\theta(x-x^{\prime})\sin(kx-kx^{\prime})\psi_{E}^{+}(x^{\prime})
\end{align*}
We see that $\psi_{E}^{+}(x)\propto\cos(kx)$ for $x\leq a$. For $a\leq x\leq
b$ (barrier region) and $E<V_{0}$, $\psi_{E}^{+}(x)$ will be a combination of
growing and decaying exponentials with decay constant $\kappa$, where
$\kappa^{2}=2m\left(  V_{0}-E\right)  $. Enforcing continuity of the wave and
its slope at $x=a$ (the inner barrier edge) leads to the equation specifying
the [below-the-barrier] energies of any even-parity, stationary quasibound
states in this `leaky' well:%
\begin{align}
1  &  =\left[  \frac{\kappa}{k}\cos(ka)+\sin(ka)\right]  \frac{2mV_{0}}%
{\kappa}%
{\displaystyle\int\limits_{a}^{b}}
\,dx\,g\left(  \sqrt{\frac{2kx}{\pi}}\right)  \exp(\kappa a-\kappa
x)\nonumber\\
&  +\left[  \frac{\kappa}{k}\cos(ka)-\sin(ka)\right]  \frac{2mV_{0}}{\kappa}%
{\displaystyle\int\limits_{a}^{b}}
\,dx\,g\left(  \sqrt{\frac{2kx}{\pi}}\right)  \exp(\kappa x-\kappa a)
\label{criterion-twin-even}%
\end{align}
The details have been omitted, since they parallel those already given in
connection with the example of the `leaky' spherical well discussed in Sec.
VII. In fact, Eq.(\ref{criterion-twin-even}) is very similar to
Eq.(\ref{criterion-leaky-well}), and can be analyzed in much the same way. In
particular, we are led to the following observations:

\begin{enumerate}
\item In the thick barrier limit ($b\rightarrow\infty$) the last integral on
the right of Eq.(\ref{criterion-twin-even}) diverges, whereas the remaining
terms are bounded for all $k,\kappa>0$. It follows that the coefficient of the
divergent integral must vanish in this limit. Careful analysis of
Eq.(\ref{criterion-twin-even}) shows that for large $b$%
\[
\frac{\kappa}{k}\cos(ka)-\sin(ka)\sim b^{3/2}\exp\left(  -\kappa w\right)
\]
Since the zeros of the expression on the left locate the energies of
even-parity states in the finite square well of width $2a$, the even-parity
stationary quasibound levels evidently merge with those bound state energies
in the limit $b\rightarrow\infty$. Also since there is but one even-parity
state at each energy -- either with or without the `sourcing' potential $V(x)$
-- the stationary quasibound state waveforms must converge to the even-parity
bound state wave functions of the finite well in this limit.

\item Unlike the odd-parity case, there is no minimum barrier thickness
required to sustain even-parity quasibound states in the central well, so at
least one such state always exists. A more refined statement follows by
writing Eq.(\ref{criterion-twin-even}) in the form%
\[
1=A\frac{\cos(ka)}{ka}-B\sin(ka)
\]
where $A$ and $B$ are the same expressions encountered previously in our study
of the `leaky' spherical well, and subject to the estimates
Eq.(\ref{bounds-AB}). The reasoning employed there indicates that the term
involving $A$ is dominant for a narrow barrier, leaving the approximate
equation%
\[
1\simeq2mV_{0}aw\frac{\sinh\left(  \kappa w\right)  }{\left(  \kappa w\right)
}\left[  2g\left(  \sqrt{\frac{2ka}{\pi}}\right)  \frac{\cos(ka)}{ka}\right]
\]
Considered as a function of $ka$, the bracketed expression clearly diverges as
$ka\rightarrow0$ (thereby guaranteeing at least one root no matter how narrow
the barrier), but diminishes rapidly and oscillates about zero with
ever-decreasing amplitude as $ka$ grows larger. Numerical investigation shows
that the first (highest) maximum is reached for $ka\simeq5.90$ and the
function value there is $\approx1/248$. We conclude that multiple even-parity,
stationary quasibound states are unsustainable if%
\begin{equation}
2mV_{0}aw\frac{\sinh\left(  \kappa w\right)  }{\left(  \kappa w\right)  }%
\leq\sqrt{2mV_{0}a^{2}}\sinh\left(  \sqrt{2mV_{0}w^{2}}\right)  \lesssim248
\label{cutoff-thin barrier-even}%
\end{equation}
Again, the bound of Eq.(\ref{cutoff-thin barrier-even}) is not rigorous, but
represents a reasoned estimate that can guide more in-depth studies.
\end{enumerate}

For resonances in the scattering from twin symmetric barriers, we turn again
to the work of Maheswari, et. al. \cite{Maheswari}, this time with focus on
the symmetric states described in their study. With $V_{0}a^{2}=72$ (in units
where $\hbar^{2}=2m=1$) and $w=0.5\,a$ they report three symmetric resonances
below the top of the barrier ($E_{r}<V_{0}$), at energies $E_{r}=0.219$,
$1.955$, and $5.298$. Using the same parameter values, we find just one
even-parity, stationary quasibound state at energy $E=0.199$. The wave
function for this quasibound state is shown in Fig. 4, where it is compared
with the bound state wave function to which it converges in the thick barrier
limit $b\rightarrow\infty$. The two are in good agreement where they must be,
viz., within the central well and near the inner barrier edge; they differ
noticeably in the outer half of the barrier and beyond. Our finding of only
one quasibound state is actually inconsistent with the bound of
Eq.(\ref{cutoff-thin barrier-even}), which for $V_{0}a^{2}=72$ predicts a
critical barrier thickness $w_{cr}\simeq0.480\,a$. In fact, the actual value
is about $7\%$ larger, at $w_{cr}\simeq0.517\,a$. This level of discrepancy is
not surprising given the lack of rigor inherent in
Eq.(\ref{cutoff-thin barrier-even}).%
\begin{figure}[ptb]%
\centering
\fbox{\includegraphics[
natheight=3.041500in,
natwidth=4.166700in,
height=2.5278in,
width=3.4134in
]%
{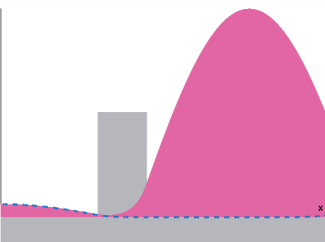}%
}\caption{The lowest-lying even-parity quasibound state in the `leaky' well
formed by twin barriers, with $V_{0}a^{2}=72$ and $w=0.5\,a$. The bound state
wave (envelope) of the well resulting from infinitely thick barriers is shown
as the dashed curve, and both are displayed against the backdrop of the twin
barrier potential $V(x)$. Only the half-axis $x\geq0$ is shown.}%
\label{figure4}%
\end{figure}

\section{Summary and Conclusions}

In this paper we have advocated the viewpoint that states described as
`quasibound' should exhibit a connectedness (in the mathematical sense) to
true bound states through the growth of some parameter. And we have
demonstrated how this connectedness can be formulated in a rigorous way using
the novel concept of quantum histories, or timelines, that span the Hilbert
space of physical states. The principle that a change of gauge cannot
influence the outcome of a measurement restricts the quasibound states to one
of two basic types: (1) stationary quasibound states that are characterized by
real energies and Schr\"{o}dinger wave functions that are everywhere bounded
(but not square-integrable), and (2), resonant states having complex energies
(and consequently divergent waveforms) that correspond to poles in the complex
$E$-plane of the scattering or $S$-matrix. Heretofore, only the latter have
been recognized in much of the quasibound state literature. That both are
rooted in the same underlying principle of connectedness leads to the
expectation that stationary quasibound states and resonant states admit
identical perturbation expansions in the growth parameter whenever such series
developments exist, i.e., the two quasibound types cannot be distinguished in
the perturbative regime.

The general theory has been applied to three diverse systems exemplifying
distinct spectral characteristics (bounds, degeneracies) of the Hamiltonian
for the embedding continuum. Within each such application class, the
stationary quasibound states of a model system have been explored and compared
with previously reported results for resonant states. The calculations are
aimed at highlighting the differences between the two basic quasibound types,
while also emphasizing their common origins. Future work will focus on further
exploring these model systems, while also entertaining other application
classes that encompass problems of widespread physical interest.

Our model results confirm that stationary quasibound states with the
properties advertised are prescribed correctly by the theory. But under what
circumstances might they actually be observed? Given that these are states
with real energies connected to true bound states by the growth of a
parameter, the Adiabatic Approximation provides an answer: starting from a
bound state, evolution proceeds through the connected quasibound state as the
controlling parameter increases, \textit{provided the growth rate is `small'}.
To formulate this idea more carefully, we follow Schiff \cite{Schiff} in
writing (approximately) the transition amplitude from the initial state
labeled $|\,n\,\rangle$ to any other instantaneous eigenstate $|\,k\,\rangle$
of the full Hamiltonian $\widehat{H}$ as%
\[
a_{k}\left(  t\right)  \approx\langle\,k\,|\frac{\partial\widehat{H}}{\partial
t}|\,n\,\rangle\frac{\exp\left(  iE_{kn}t\right)  -1}{iE_{kn}^{2}},
\]
where $E_{kn}\equiv E_{k}-E_{n}$. We observe with Schiff that the probability
of populating states other than the initial one simply oscillates in time, and
shows no steady increase over long periods. But the amplitude of these
oscillations is fixed by $\left\vert E_{kn}^{-1}\langle\,k\,|\partial
\widehat{H}/\partial t|\,n\,\rangle\right\vert $, i.e., directly proportional
to the growth rate and inversely proportional to the level spacing. For
continuous spectra ($E_{kn}\rightarrow0$), negligible transition amplitude
would require infinitesimal growth rates, but the continuum approximation is a
mathematical convenience that is never realized in practice. Confining
boundaries introduce level separations and matrix elements that scale
inversely with system size. If the length scale of the system is set by $L$,
we can expect $E_{kn}\sim L^{-2}$ and $|\,k\,\rangle\sim$ $L^{-1/2}$. In such
systems $a_{k}\left(  t\right)  $ can be kept small if $\partial
\widehat{H}/\partial t\sim L^{-1}$. We conclude that experimental evidence for
the ideas presented here is likely to be found among mesoscopic systems where
the interaction can be varied slowly and with great precision.


\begin{thebibliography}{99}                                                                                               %


\bibitem {Schrodinger}E. Schr\"{o}dinger, Ann. Phys. (Leipzig) \textbf{80},
437 (1926).

\bibitem {Oppenheimer}J. R. Oppenheimer, Phys. Rev. \textbf{31}, 66 (1928).

\bibitem {Gamow}G. Gamow, Z. Phys. \textbf{51}, 204 (1928).

\bibitem {Fowler}R.H. Fowler and L. Nordheim, Proc. Roy. Soc. \textbf{A 119},
180 (1928).

\bibitem {UsageNote}Resonances always connote decaying states, described by
the stationary form $\psi_{E}\exp\left(  -iEt/\hbar\right)  $ with complex
energy $E$ whose imaginary part directly determines the state lifetime. Common
usage suggests that quasibound states are synonomous with resonances, but a
distinction often is drawn when they are constructed following a procedure
that endows them with real energies. In our view, `quasibound' is the more
inclusive term, and our usage in this manuscript reflects that bias.

\bibitem {Harrell}E. M. Harrell II, Proceedings of \ Symposia in Pure
Mathematics, \textbf{76.1}, 227 (2007).

\bibitem {Gunapala}S.D. Gunapala and S.V. Bandara, Microelectron. J.
\textbf{30}, 1057 (1999).

\bibitem {Razeghi}M. Razeghi, Microelectronics Journal \textbf{30}, 1019 (1999).

\bibitem {Lee}C.D. Lee, S.K. Noh, and Kyu-Seok Lee, Superlattices and
Microstructures \textbf{21}, 101 (1997).

\bibitem {Price}P.J. Price, Microelectron. J. \textbf{30}, 925 (1999).

\bibitem {Anemogiannis}E. Anemogiannis, E.N. Glytsis, and T.K. Gaylord,
Microelectron. J. \textbf{30}, 935 (1999).

\bibitem {Rakityansky}S.A. Rakityansky, Phys. Rev. B \textbf{68}, 195320 (2003).

\bibitem {Rihani}Samir Rihani, Hideaki Page, and Harvey E. Beere,
Superlattices and Microstructures \textbf{47}, 288 (2010).

\bibitem {Moyer}Curt A. Moyer, J. Phys. A \textbf{30}, 7537 (1997). The
generalized Airy functions cited in this work are related to the Scorer
functions introduced in Sec. VI of the present paper as $e_{0}(z)\equiv\pi
Hi(-z)$ and $E_{0}(z)\equiv-\pi Gi(-z)$.

\bibitem {Moyer4}C.A. Moyer, arXiv:1305.5525v1.

\bibitem {Messiah}A. Messiah, \textit{Quantum Mechanics Vol. II}, (John Wiley
\& Sons, New York, 1966), p. 713.

\bibitem {Davydov}A.S. Davydov, \textit{Quantum Mechanics}, 2nd ed. (Pergamon
Press, New York, 1965), p. 542.

\bibitem {Landau}L.D. Landau and \ E.M. Lifshitz, \textit{Quantum Mechanics},
2nd ed. (Pergamon Press, Oxford, 1965), p. 269.

\bibitem {Scorer}R.S. Scorer, Quart. J. Mech. Appl. Math. \textbf{3}, 107 (1950).

\bibitem {Moyer2}Curt A. Moyer, J. Phys. C \textbf{6}, 1461 (1973).

\bibitem {Moyer3}C.A. Moyer, Ph.D. Thesis, State University of New York at
Stony Brook, 1971, p. \ (unpublished).

\bibitem {Ludviksson}A. Ludviksson, J. Phys. A \textbf{20}, 4733 (1987).

\bibitem {Cavalcanti}R.M. Cavalcanti, P. Giacconi and R. Soldati, J. Phys. A
\textbf{36} 12065 (2003).

\bibitem {Alvarez}G. Alvarez and B. Sundaram, Phys. Rev. A \textbf{68}, 013407 (2003).

\bibitem {Abramowitz2}\textit{Handbook of Mathematical Functions} edited by M.
Abramowitz and I. A. Stegun, (Dover, New York, 1965), p. 364.

\bibitem {Merzbacher}E. Merzbacher, \textit{Quantum Mechanics}, 2nd ed. (John
Wiley \& Sons, New York, 1970), p. 200-2.

\bibitem {Erdelyi}A. Erd\'{e}lyi, \textit{Asymptotic Expansions} (Dover
Publications Inc., New York, 1956).

\bibitem {Abramowitz4}\textit{Handbook of Mathematical Functions} edited by M.
Abramowitz and I. A. Stegun, (Dover, New York, 1965), p. 300. The integral
representation for $g(\ldots)$ reported here follows directly from its
definition in terms of the Fresnel sine and cosine integrals.

\bibitem {Maheswari}A.U. Maheswari, P. Prema, S. Mahadevan and C.S. Shastry,
Pramana $-$ J. Phys. \textbf{73}, 969 (2009).

\bibitem {Schiff}L.I. Schiff, \textit{Quantum Mechanics}, 3rd ed. (McGraw-Hill
Inc., New York, 1968), p. 291.
\end{thebibliography}
\end{document}